\documentclass[10pt]{article}

\pdfoutput=1

\usepackage{amsmath,amssymb,amsfonts,amscd,mathrsfs}
\usepackage{xcolor}
\definecolor{darkblue}{rgb}{0.1,0.1,.7}
\usepackage[colorlinks, linkcolor=darkblue, citecolor=darkblue, urlcolor=darkblue, linktocpage]{hyperref} 
\usepackage[square, comma, compress,numbers]{natbib}
\usepackage[]{graphicx}
\usepackage{geometry}
\usepackage[margin=10pt,font=small,labelfont=bf]{caption}
\usepackage{ifthen}
\usepackage{tikz}
\usepackage{subcaption}
\usepackage{booktabs,multirow}
\usepackage{hhline}
\usepackage{tablefootnote}

\usepackage{dsfont} 

\usepackage{titlesec}
\titleformat*{\section}{\large\bfseries}
\titleformat*{\subsection}{\normalsize\bfseries}
\titleformat*{\subsubsection}{\normalsize\it}
\titleformat*{\paragraph}{\normalsize\bfseries}
\titleformat*{\subparagraph}{\normalsize\bfseries}

\newcommand{\reef}[1]{(\ref{#1})}
\def\vareps{\varepsilon}

\def\eps{\epsilon}

\def\AF{\ensuremath{_{\rm AF}}}
\newcommand{\beq}{\begin{equation}} 
\newcommand{\eeq}{\end{equation}}
 
\def\nn{\nonumber} 
\def\
 {\mathbb{Z}} 
\def\bR {\mathbb{R}} 
\def\bC {\mathbb{C}} 
\def\calO {{\cal O}}

\def\calN {{\cal N}} 
\def\calM {{\cal M}}

\def\calT {{\cal T}} 
\def\calC {{\cal C}} 
\def\calR {{\cal R}} 
\def \barcalC {\overline{\calC}}
\def \cC {$\overline{\calC}$}

\def\bZ {\mathbb{Z}} 
\def\bC {\mathbb{C}} 
 
\def\half{{\textstyle\frac 12}}

\def\ge{\geqslant}
\def\le{\leqslant}

\def\<{\langle}
\def\>{\rangle}
\newcommand{\diffop}[2]{\ifthenelse{\equal{#2}{1}}{\frac{\mrm{d}}{\mrm{d} #1}}{\frac{\mrm{d}^#2}{\mrm{d} #1^#2}}}

\newcommand{\be}{\begin{equation}}
\newcommand{\ee}{\end{equation}}
\newcommand{\bea}{\begin{eqnarray}}
\newcommand{\eea}{\end{eqnarray}}

\newcommand{\mrm}[1]{{\mathrm #1}}

\def\kappa{\varkappa} 
\def\tr{\mathrm{tr}}
\usepackage[normalem]{ulem}

\usepackage{accents}
\newlength{\dhatheight}

\numberwithin{equation}{section}
\usepackage[tocgraduated]{tocstyle}

\begin{document}

\vspace*{-.6in} \thispagestyle{empty}
\begin{flushright}
\end{flushright}
\vspace{1cm} {\large
\begin{center}
{\bf Walking, Weak first-order transitions,
and Complex CFTs}
\end{center}}
\vspace{1cm}
\begin{center}
{\bf Victor Gorbenko$^a$, Slava Rychkov$^{b,c}$,  Bernardo Zan$^{d,b,c}$}\\[2cm] 
{
$^{a}$ Stanford Institute for Theoretical Physics, Stanford University, Stanford, CA 94305, USA\\
$^b$  Institut des Hautes \'Etudes Scientifiques, Bures-sur-Yvette, France\\
$^c$  Laboratoire de physique th\'eorique, D\'epartement de physique de l'ENS \\
\'Ecole normale sup\'erieure, PSL University,
Sorbonne Universit\'es, \\
UPMC Univ.~Paris 06, CNRS, 75005 Paris, France
\\
$^d$ Institut de Th\'eorie des Ph\'enom\`enes Physiques, EPFL, CH-1015 Lausanne, Switzerland}
\vspace{1cm}
\end{center}

\vspace{4mm}

\begin{abstract}
	We discuss walking behavior in gauge theories and weak first-order phase transitions in statistical physics.
Despite appearing in very different systems (QCD below the conformal window, the Potts model, deconfined criticality) these two phenomena both imply approximate scale invariance in a range of energies and have the same RG interpretation: a flow passing between pairs of fixed point at complex coupling. We discuss what distinguishes a real theory from a complex theory and call these fixed points complex CFTs. By using conformal perturbation theory we show how observables of the walking theory are computable by perturbing the complex CFTs. This paper discusses the general mechanism while a companion paper \cite{part2} will treat a specific and computable example: the two-dimensional $Q$-state Potts model with $Q>4$. Concerning walking in 4d gauge theories, we also comment on the (un)likelihood of the light pseudo-dilaton, and on non-minimal scenarios of the conformal window termination.
	
 \end{abstract}
\vspace{.2in}
\vspace{.3in}
\hspace{0.7cm} July 2018

\newpage

\setcounter{tocdepth}{3}

{
\tableofcontents
}

\section{Introduction}

Walking is a somewhat mysterious behavior which can conjecturally be exhibited by some four-dimensional (4d) gauge theories. 
In a walking gauge theory, the gauge coupling is supposed to run slowly at intermediate energies, where the theory is approximately scale invariant, while at low energies the coupling starts running fast again, leading to confinement and chiral symmetry breaking. Originally this has been dreamed of in the context of technicolor scenarios of electroweak symmetry breaking \cite{Holdom:1981rm,Yamawaki:1985zg,Appelquist:1986an}. 

A number of curious opinions about walking can be found in the literature. Walking is supposed to happen just below the end of the conformal window \cite{Kaplan:2009kr}. It is believed by some that walking theories contain a naturally light pseudo-dilation in the spectrum \cite{Yamawaki:1985zg}. There are doubts if walking may naturally occur in theories with a small number of colors \cite{Luty:2008vs}. {We warn the reader that only the first of these three opinions will find a confirmation in our analysis. The above definition of walking itself also needs revision, since as we will see it's not the gauge coupling which walks. We collected here this mix of opinions to stress that, at least to us, walking appears a rather controversial subject where much confusion lingers.} This is also due to the fact that probing this scenario directly by lattice Monte Carlo simulations remains a hard task. 

In this paper, we will first improve understanding of walking by drawing intuition from a much simpler example of this behavior, belonging to the realm of statistical physics: the $Q$-state Potts model in 2d. This model is known to have a conformal phase for $Q<4$, and a first-order phase transition at $Q>4$. For $Q\gtrsim 4$, the transition is \emph{weakly} first-order: 
the correlation length is much larger than the lattice spacing. This was understood by statistical physicists in the 1980s \cite{NauenbergPRL,CNS} in terms almost identical to walking (one difference being that the Potts model has a strongly relevant singlet scalar whose coefficient is tuned to zero to reach the transition).  As far as we know, the connection between walking and weakly first-order phase transitions is being made here for the first time in the high energy physics literature.\footnote{In condensed matter/statistical physics this connection is not forgotten, as we will see in section \ref{sec:further}. Walking is one of two known mechanisms which can explain weakness of a first-order phase transition, the other one being tuning, see section \ref{sec:Tuning} and appendix \ref{sec:1st-order}.}

Our second goal is to demystify the \emph{fixed points at complex coupling}, often invoked in discussions of walking. We will formalize these fixed points as \emph{complex conformal field theories (CFTs)}, a concept that we introduce. Complex CFTs are non-unitary, but they are sufficiently different from other commonly occurring non-unitary CFTs that they deserve a separate name. For example, 2d complex CFTs have a complex central charge $c$. In spite of this and other unusual features, we will argue that complex CFTs are nonperturbatively well defined. We will discuss, in general terms, how this new language can be used to describe some aspects of walking.

The paper is structured as follows. In section \ref{sec:hierarchy} we present walking from renormalization group (RG) point of view: as a general mechanism for generating hierarchies in quantum field theory (QFT). We also briefly review a more common mechanism known as tuning. In the same section we give a first introduction to the concept of complex CFTs.

Sections \ref{sec:statphys},\ref{sec:BZ},\ref{sec:further} focus on concrete systems exhibiting walking. In section \ref{sec:statphys} we discuss how walking is realized in the 2d $Q$-state Potts model, including a detailed introduction to this lattice model for the benefit of high energy physicists.
Section \ref{sec:statphys} discusses various aspects of walking in 4d gauge theories. In particular, we explain why we don't believe in the parametrically light pseudo-dilaton. Finally in section \ref{sec:further} we discuss a recent example of walking that emerged in condensed matter physics, in the context of ``deconfined criticality".

Section \ref{sec:Complex} is devoted to complex CFTs. We build upon intuitive understanding of the difference between RG flows in the space of real vs complexified couplings, towards a more formal definition of the concept of a complex QFT and a complex CFT. We explain how the real vs complex classification differs from the more familiar unitary vs non-unitary classification. In particular we give examples of non-unitary but real theories. Finally we come back to the connection between complex CFTs and walking. We present a computational paradigm, a kind of conformal perturbation theory, which allows to compute certain properties of walking RG flows in terms of CFT data of complex CFTs. In this paper we only discuss general features of this paradigm. In companion paper \cite{part2} we will show its usefulness by studying the walking behavior of the 2d Potts model at $Q>4$. We will see that it allows for many concrete applications, tests, and predictions. 

In section \ref{sec:conclusions} we conclude. The paper has several appendices. Appendix \ref{sec:1st-order} reminds that not all weakly first-order phase transitions are explainable by walking, some being due to tuning. Appendix \ref{sec:BKT} explains the difference between the physics of walking and the BKT transition. Appendix \ref{sec:furtherPotts} contains further details about the Potts model, in particular in $d>2$. Appendix \ref{sec:largeN} discusses features of conformal window and walking in 4d gauge theories arising in the large $N$ limit.

\section{Walking as a mechanism for hierarchy}
\label{sec:hierarchy}

\def\LUV{\Lambda_{\rm UV}}
\def\LIR{\Lambda_{\rm IR}}
\def\xUV{\ell_{\rm UV}}
\def\xIR{\ell_{\rm IR}}

This section will define walking using the language of RG, without specializing to any particular microscopic description. Walking is one of two known robust mechanisms for generating hierarchy in QFT, the other one being the much more familiar `tuning'. The question of hierarchies being of extreme importance, this explains why one should a priori be interested in walking.

Hierarchy is a separation of scales. A hierarchy in quantum field theory means that the theory contains two distance scales $\xUV\ll\xIR$ (or equivalently two energy scales $\LUV\gg\LIR$), the physics being approximately scale invariant in the intermediate range between them. The scale $\xUV$ can be thought of as a short-distance cutoff. The scale $\xIR$ in high energy physics is usually related to the inverse mass of some particle, while in statistical physics it is the correlation length.

Hierarchies are a familiar feature of theories with a logarithmically running coupling, such as the usual QCD. Although the coupling runs slowly, and one may be tempted to say poetically that it `walks',\footnote{Frank Wilczek used to say ``You must walk before you run!" in his colloquia, referring to the QCD gauge coupling.} in our technical classification this is actually an example of a (mild) tuning and not of walking, see below.

\subsection{Tuning}
\label{sec:Tuning}

Tuning mechanism for hierarchies is completely standard and utterly familiar to QFT practitioners, but let's review it anyway to set the stage. From many available prior discussions, ours will stay closest to \cite{Luty:2004ye, Rattazzi:2008pe}.
 
In this mechanism a hierarchy results from the fact that an RG trajectory describing the QFT starts close to a CFT and remains close to it for a long time. We can thus think of the RG flow in terms of the perturbing operators added to the CFT. For the flow to stay close to the CFT, we should worry in particular about the coefficients of all \emph{relevant} perturbations, which must be assumed small. 

Assuming for simplicity that there is just one relevant operator, the arising hierarchy is controlled by the size of its coefficient. At some UV scale where the microscopic theory is matched onto the CFT plus a perturbation, the theory is described by
\beq
\label{eq:CFTtuning}
{\rm CFT}+c\,\Lambda_{\rm UV}^{d-\Delta}\int d^dx\, \calO_\Delta(x)\,,
\eeq
where $\calO_\Delta$ is a scalar operator of scaling dimension $\Delta<d$, and $c\ll 1$.\footnote{By $d$ we denote the full number of dimensions, which includes time if we work in Minkowski signature.} The necessity to take $c\ll 1$ is why this scenario is called ``tuning".
Then assuming the coupling $c$ does not flow to a fixed point, the relation between the UV scale and the IR scale, at which the departure from the CFT becomes significant, is\footnote{We don't keep track of factors of $4\pi$, which would be useful in practical applications of this sort of naive dimensional analysis.}
\beq
\label{eq:hier}
\LIR\sim c^{\frac {1}{d-\Delta}}\LUV\,.
\eeq

In applications of this scenario in high energy physics, there is, justifiably, much preoccupation with how ``natural" the implied tuning is. If the operator $\calO_\Delta$ transforms non-trivially under some global symmetry present in the CFT, the assumption of a small coefficient $c$ is considered ``technically natural" in QFT jargon, because $c=0$ would be preserved by RG evolution. Put another way, the smallness of this coefficient can be explained by requiring that the symmetry be approximately preserved in the microscopic description of our theory. This is just 't Hooft's naturalness criterion \cite{tHooft:1979rat} restated in the CFT language. 

A more problematic case is when $\calO_\Delta$ is a full singlet of the CFT global symmetry group. In this case a fully natural hierarchy is never possible. However there is a way to turn a \emph{mild tuning} into a large hierarchy, provided that $\calO_\Delta$ is \emph{weakly relevant}, that is if $d-\Delta\ll 1$ \cite{Luty:2004ye}. To see this, notice that if both $c$ and $d-\Delta$ are somewhat small, say 0.1, then Eq.~\reef{eq:hier} predicts the hierarchy $\LUV/\LIR\sim 10^{10}$. 

The above-mentioned QCD example can be seen as a limiting case of the latter situation when $\Delta\to d$ and the operator is marginally relevant. In this case the relation between the IR scale ($\Lambda_{\rm QCD}$) and the UV cutoff is exponential in the inverse of the bare gauge coupling. But to enjoy this exponential hierarchy, we must still assume that the gauge coupling is somewhat small at the cutoff, hence mild tuning.

In condensed matter/statistical physics context, the tuning mechanism explains the weakness of some first-order phase transitions, see appendix \ref{sec:1st-order}.

\subsection{Walking} 
\label{sec:walking}

We will now discuss walking which is our main interest. In this case the CFT picture is a bit more complicated, and it is convenient to present first a more intuitive picture based on the RG. We consider an RG flow of a coupling $\lambda$, of unspecified origin, and a singlet under the global symmetry. We assume that the beta-function takes near $\lambda=0$ the form ($t=\log E$)
\beq
\beta(\lambda)=\frac{d\lambda}{dt}=-y-\lambda^2 +O(\lambda^3) \,,
\label{eq:RGwalk}
\eeq
where $y$ is a small parameter, and the higher order terms are assumed to have $O(1)$ coefficients so that Eq.~\reef{eq:RGwalk} is trustworthy at $|\lambda|\lesssim 1$. While the choice of $\lambda=0$ may seem special, there is no loss of generality here, as we can first assume that the 
beta-function takes this form with $\lambda-\lambda_0$ instead of $\lambda$ in the r.h.s., and eliminate $\lambda_0$ by a shift.\footnote{Even more generally the $\lambda^2$ term could be $A(\lambda-\lambda_0-By )^2$, and we can set $A\to 1$, $\lambda_0\to0$, $B\to 0$ shifting and rescaling $\lambda$.} 

Of course describing the flow in terms of just one coupling is an idealization. What we imagine is that all other couplings characterizing the flow are irrelevant, and so their effect on the flow of $\lambda$ can be neglected.\footnote{\label{note:rel}It is also possible that a few of them are relevant and then have to be finetuned small. This is what happens in the Potts model where one has to finetune the temperature, see section \ref{sec:2dPottsWalk}.}

\begin{figure}[htbp] 
	\centering
	\begin{subfigure}{.45\textwidth}
		\centering
		\includegraphics[width=.6\linewidth]{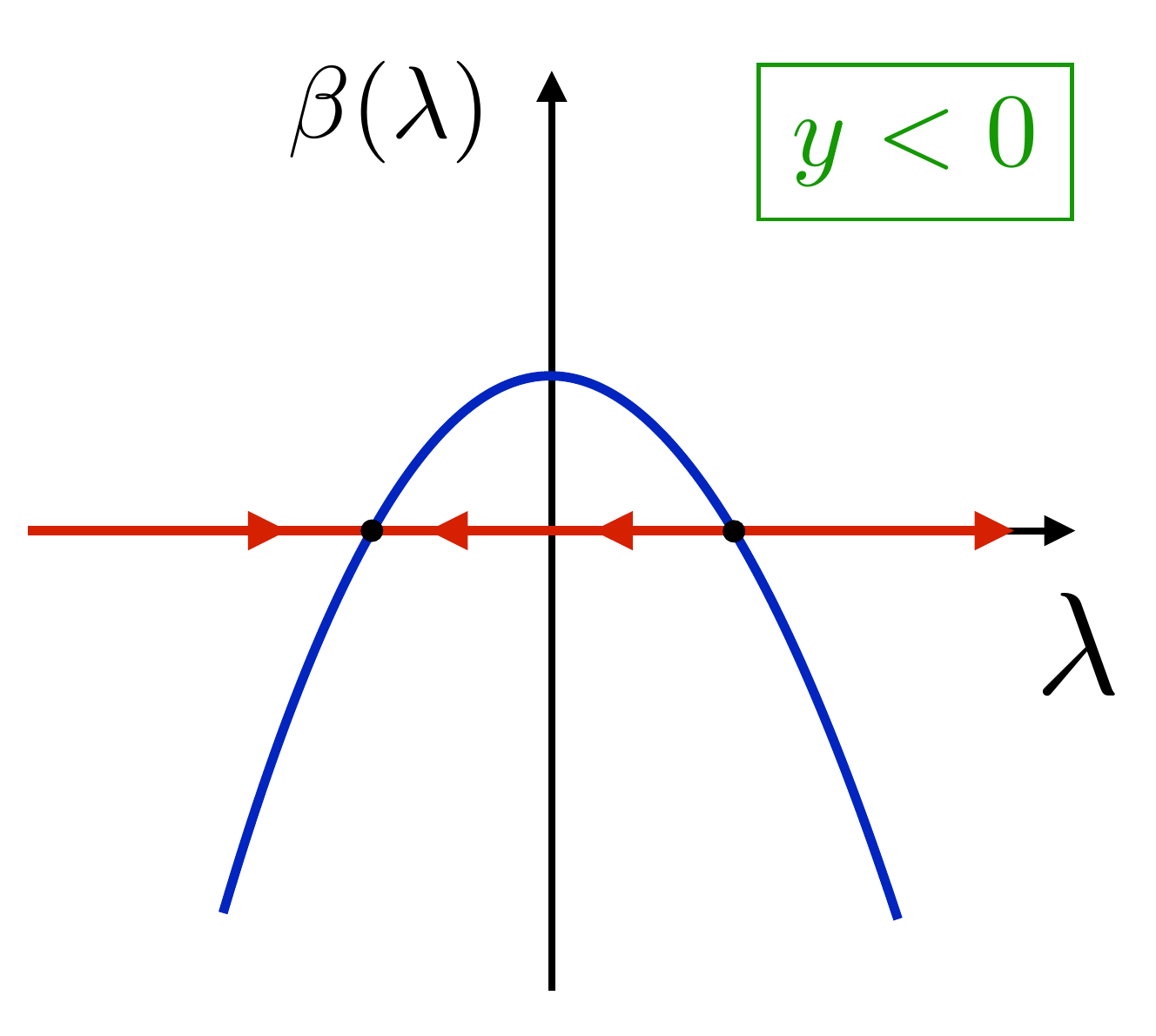}
	\end{subfigure}%
	\begin{subfigure}{.45\textwidth}
		\centering
		\includegraphics[width=.6\linewidth]{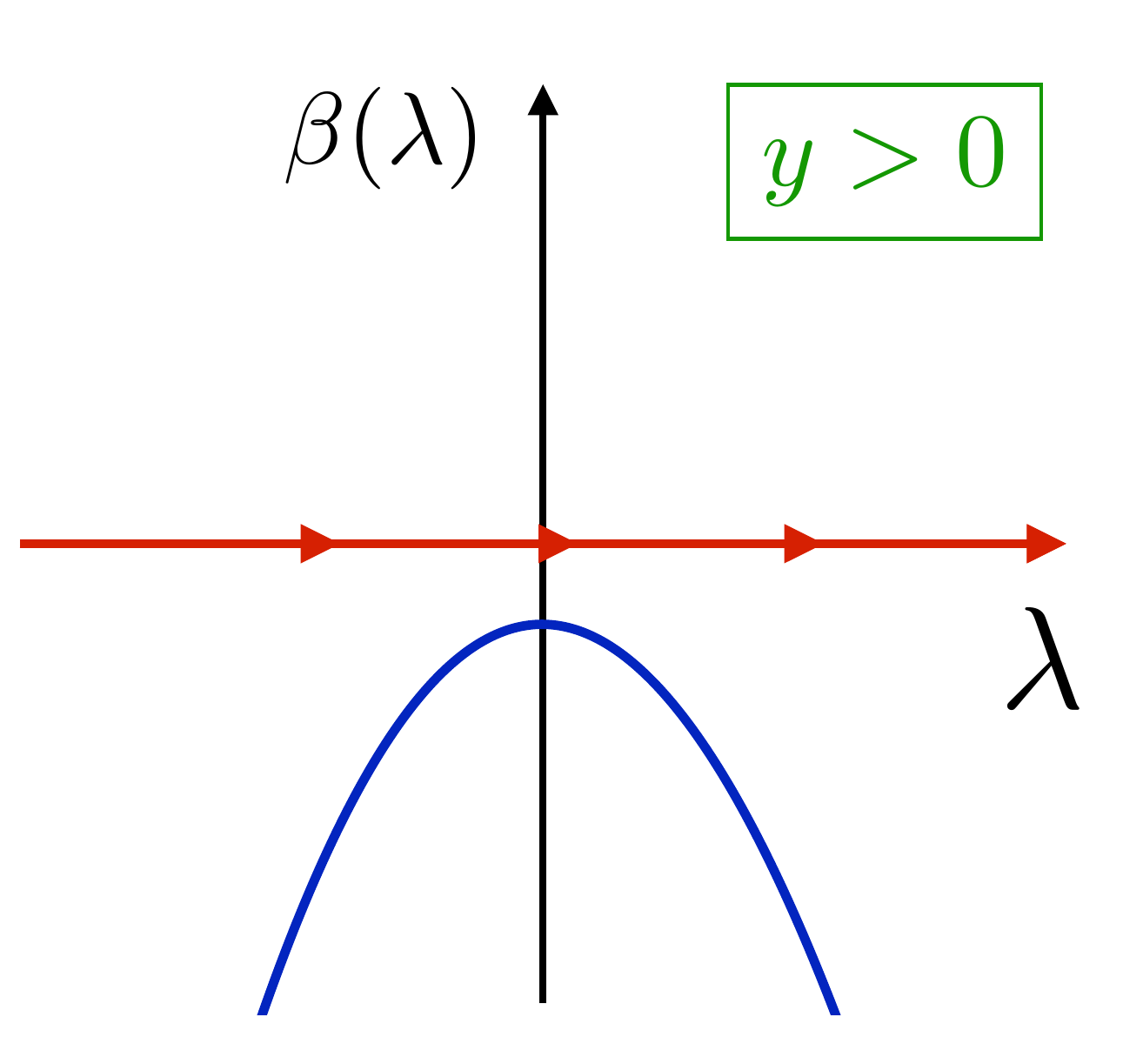}
	\end{subfigure}
	\caption{Structure of RG flow for real coupling for $y<0$ (left) and $y>0$ (right). 
	}
	\label{fig:fig1}
\end{figure}

Assuming that the coupling $\lambda$ is real, physics described by the beta-function \reef{eq:RGwalk} is very different depending on the sign of $y$. Suppose first that $y<0$. Then we have two real fixed points $\lambda_{\pm}=\pm \sqrt{|y|}$ (see Fig.~\ref{fig:fig1}). The $\lambda_+$ fixed point is a UV fixed point in the sense that it cannot be reached by flowing from short distances. The $\lambda_-$ fixed point is an IR fixed point as it can be reached flowing both from the UV fixed point and also from large negative $\lambda$, provided that in this range the microscopic description happens to match approximately the RG flow described by the above beta-function.

Concerning the CFTs describing these fixed points, the operator $\calO_{\lambda}$ to which $\lambda$ couples will have dimension
\beq
\label{eq:Dpm}
\Delta_{\pm}=d+\beta'(\lambda_\pm)\approx d\mp 2\sqrt{|y|}\,.
\eeq
For $|y|\ll 1$ this dimension is weakly relevant at the $\lambda_+$ fixed point. This CFT can be used to realize the mildly tuned hierarchy scenario of the previous section, flowing out in the positive $\lambda$ direction.
 
Suppose instead that $y>0$. Then there is no fixed point at real $\lambda$, at least not within the region of validity of the assumed approximate beta-function. A flow starting at $\lambda\sim -1$ will eventually go through to $\lambda\sim1$, but for small $y$ it will slow down and linger around $\lambda\sim 0$. 
How much the flow lingers can be estimated by computing the RG time of passage:
\beq
\Delta t \sim - \int_{-1}^1 \frac{d\lambda}{\beta(\lambda)} \sim  \frac{\pi}{\sqrt{y}}\,.
\eeq
One can also use the exact solution of the beta-function equation neglecting $O(\lambda^3)$ terms:
	\beq
	\lambda(t)=-\sqrt{y}\tan [\sqrt{y}(t-t_0)]\,.
	\eeq
If we call $\LUV$ the scale where the flow emerges from a microscopic description at $\lambda\sim-1$, and $\LIR$ the scale where the flow plunges into the unknown at $\lambda\sim 1$, we obtain:
\beq
\LUV/\LIR \sim e^{\Delta t} \sim \exp({\pi}/\sqrt{y})\,,
\label{eq:xiWalk}
\eeq
a huge ratio of scales if $y$ is small. We will refer to \reef{eq:xiWalk} as `walking scaling'.\footnote{Ref.~\cite{Kaplan:2009kr} refers to the functional form of this equation as `BKT scaling', since this is also the form of the correlation length in Berezinskii-Kosterlitz-Thouless (BKT) transition. We review the physics of BKT transition in appendix \ref{sec:BKT}. In our opinion, there are more differences than similarities between BKT transition and walking, and so we propose to avoid terminology `BKT scaling' when discussing walking in this paper and in the future.}

Given the hierarchy, we expect that the flow in the intermediate range of energies should be approximately scale invariant, so it should be close to a CFT. But to which CFT? For $y<0$ we had two CFTs, but for $y>0$ there are no CFTs in sight. 

One way out is to argue that the flow remains close to the CFT which describes the $y=0$, $\lambda=0$ point, where the fixed points join and disappear.\footnote{This point of view would be close to \cite{NauenbergPRL,CNS} where the walking scenario was first elucidated, see section \ref{sec:2dPottsWalk}.} This proposal is certainly viable and physically reasonable, and it allows to compute some quantities characterizing the flow at $y>0$, expanding around the $y=0$ point. However, there are some puzzling features with this way of thinking and computing.

One puzzle is what to do when the global symmetry of the problem depends on $y$. In the concrete examples of the Potts model and of the 4d gauge theories, the global symmetry will be $S_Q$ and $SU(N_f)\times SU(N_f) \times U(1)$ respectively, with $Q$ and $N_f$ continuous functions of $y$. Certainly there are limitations for expanding a theory with, say, $S_5$ symmetry around a theory with $S_4$ symmetry, and yet in the above proposal that's what we would have to do.

{Another puzzle is that the above discussion does not have a built-in criterion for determining the range of validity of the obtained expansions. One might think that it is $|y|\ll 1$, but this is too naive and can't be true 
because $y$ is just an arbitrary parameter, not a physically significant quantity. And indeed the naive criterion with $y=Q-4$ is violated by the 2d Potts model (see section \ref{sec:lessons}). We need a better criterion.} 

\subsubsection{Introducing complex CFTs}
\label{sec:introcomplex}
\begin{figure}[htbp] 
   \centering
   \includegraphics[width=10cm]{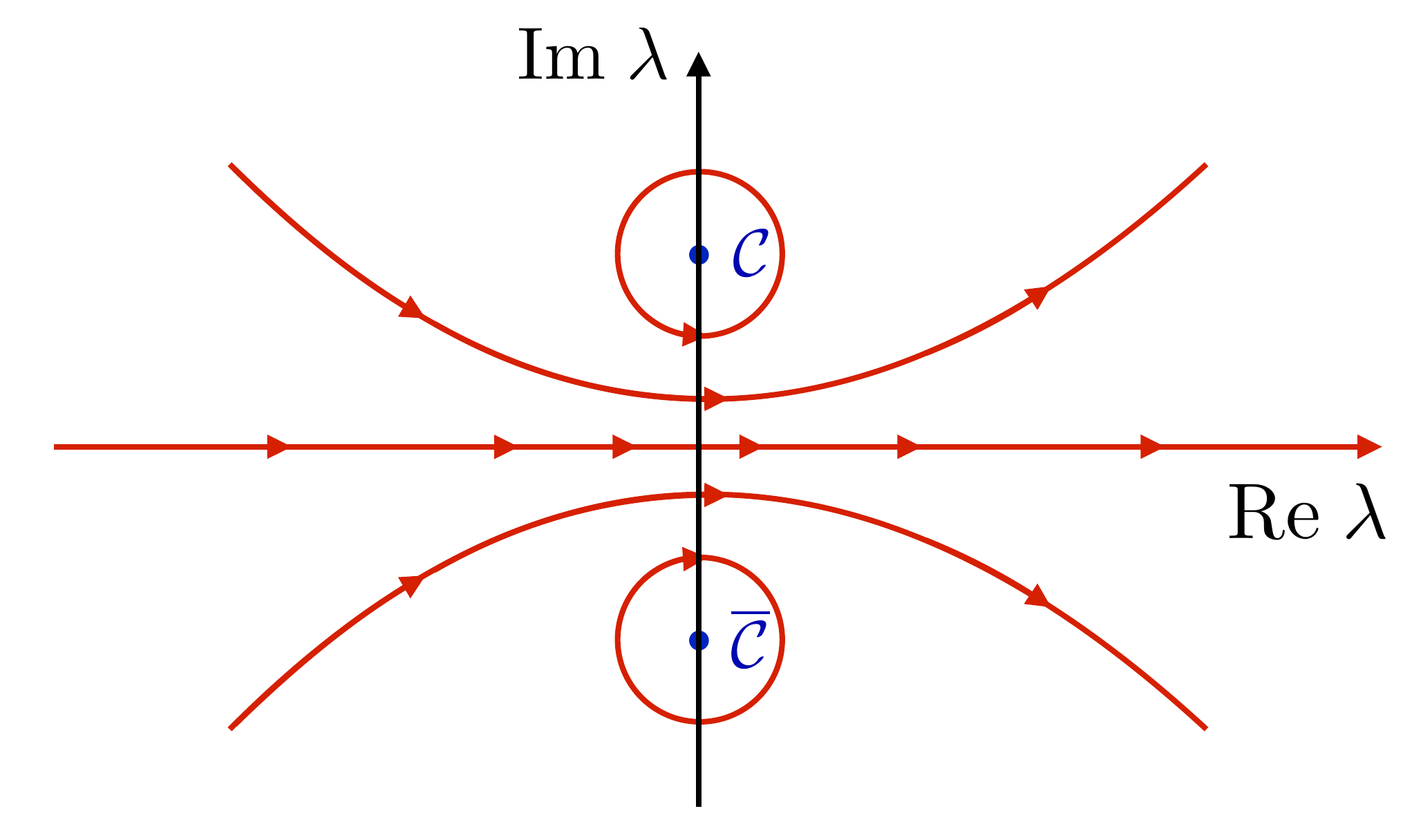} 
   \caption{Structure of RG flow in the complex coupling plane, in the approximation of dropping the higher order terms in \reef{eq:RGwalk}. Notice that including those terms will generically change the nature of RG flow trajectories around $\calC$ and $\overline\calC$, since the RG eigenvalue will then acquire a small real part $O(y^2)$, making the flow spirally in- or unwinding. See \cite{part2} for an example.} 
   \label{fig:fig2}
\end{figure}

To achieve some peace with the above puzzles, let us reconsider the fate of the fixed points at $y>0$. Of course the fixed points don't just disappear completely, but they \emph{go to the complex plane}, see Fig.~\ref{fig:fig2}. While this is often said, as far as we know until now there has not been any concrete attempt to assign physical meaning to these complex fixed points. This is precisely what we would like to do. We posit that these fixed points should be viewed as nonperturbatively defined non-unitary CFTs of a novel type, which we call \emph{complex CFTs}. To the pair of complex conjugate fixed points there will correspond a pair of complex conjugate CFTs, called $\calC$ and $\overline\calC$.

We will argue that these complex CFTs control the walking flow in the same way as the CFT appearing in \reef{eq:CFTtuning} controls the tuned flow. It is around them that one should more properly expand the flow, and not around the CFT at 
$y=0$. Doing so we readily resolve the
 first
 puzzle, 
 since $\calC$ and $\overline\calC$, living at the same value of the $y$ parameter, have the same global symmetry as the physical RG flow along the real axis.

Having recourse to $\calC$ and $\overline\calC$ also allows to determine the criteria for the walking behavior more sharply. 
For small $y>0$ let us compute the fixed point dimension $\Delta$ of the CFT operator $\calO_{\lambda}$ to which $\lambda$ couples. The fixed points being at $\lambda_*=\pm i\sqrt{y}$, we get similarly to \reef{eq:Dpm}
\beq
\Delta=d+\beta'(\lambda_*)=d\mp 2i\sqrt{y}+O(y^2)\,. 
\label{eq:delta_y}
\eeq
Notice that the dimension is complex (and complex conjugate for $\calC$ and $\overline\calC$), which will be a hallmark of complex CFTs. Notice as well that $\Delta$ is close to marginality, with the leading deviation imaginary and small.
As we will see, it is this smallness of the imaginary part of the near-marginal operator dimension 
which is necessary for walking, and not the smallness of $y$ by itself. 
Suppose then that we have some nonperturbative access to ${\rm Im}\, \Delta$, for example because we solved $\calC$ and $\overline\calC$. Then there is no need to expand around $y=0$. Instead, using the nonperturbative solution, we can simply determine the range of $y$ for which $| {\rm Im}\, \Delta |\ll 1$ holds.
This statement will be justified by means of conformal perturbation theory in section \ref{ssec:complexCFTS_walking} and in \cite{part2}.

\subsubsection{Naturalness of walking}
\label{sec:NatWalk}

As for the tuning scenario, we would like to make an assessment of how natural the walking mechanism is.

We have seen that walking needs a complex CFT $\calC$ (and its complex conjugate $\overline\calC$) with an operator whose dimension has the real part close to marginality ($d$) and the imaginary part small. An assumption of having such a CFT at our disposal certainly represents some ``finetuning in theory space", just like the assumption of having a CFT with a weakly relevant deformation in the mild tuning scenario. In fact we have seen that both these assumptions can be realized within a one-parameter family of RG flows, close to a special parameter value where a UV and an IR fixed points collide, and on two opposite sides of this value.

However, apart from this theory space finetuning, there is no further coupling finetuning required in the walking scenario (provided that all other couplings but $\lambda$ are irrelevant). Indeed, we can start the flow \emph{anywhere} on the negative real axis of $\lambda$, which represents 50\% of the a priori possible initial coupling values. It will then inexorably be drawn to $\lambda$ near 0. Basically, the flow is forced to pass between the `pillars of Hercules' $\calC$ and $\overline\calC$, because it has nowhere else to go. 

This situation can be contrasted with the mild tuning scenario, where we had to buy \emph{both} finetuning in theory space, and a mild coupling finetuning $c\ll 1$. From this point of view, walking seems less finetuned than mild tuning.

The just-given discussion only considered the naturalness of the basic walking scenario associated with the running of $\lambda$. There is an additional finetuning price if relevant singlet operators are present, whose coupling has to be tuned to zero, as for the Potts model.

\section{Walking in statistical physics: 2d Potts model}
\label{sec:statphys}

Although we borrowed the term `walking' from the physics of 4d gauge theories, 
historically the first example of walking has been observed in the 2d $Q$-state Potts model. This model is one of best known models of statistical physics, see \cite{Wu,Jacobsen2012} for reviews, but it's not as widely known to high energy physicists as it deserves. We therefore start with a mini-review. Generalizations to $d>2$ will be discussed in appendix \ref{sec:Pottsd>2}. 

\subsection{Spin and cluster definitions}
Consider a square lattice in 2d (other lattices are also possible). 
It will be important that the Potts model has two lattice definitions: either as a model of random spins living on lattice sites, or as a model random clusters, that is connected sets of lattice bonds. The two definitions agree for integer $Q\ge 2$, with the second definition providing an analytic continuation to non-integer $Q$'s.

In the spin definition, we put on every lattice site $i$ a discrete variable $s_i\in\{1,2,\ldots,Q\}$. The partition function is the sum over spin configurations:
\beq
Z_{\rm spin}=\sum_{\{s_i\}}e^{-H[\{s_i\}]}\,,
\eeq
with the lattice Hamiltonian (we include $\beta=1/T$ into the Hamiltonian) being the sum of nearest-neighbor interaction terms which energetically prefer for the spins to be identical (called the ferromagnetic case):
\beq
\label{pottsH}
H[\{s_i\}]= - \beta \sum_{\langle i j\rangle} \delta_{s_i,s_j}\,.
\eeq
For $Q=2$ this reduces, up to a constant shift, to the Ising model Hamiltonian. This model has a discrete global symmetry $S_Q$ (the permutation group). 

\begin{figure}[htbp] 
	\centering
	\includegraphics[scale=.3]{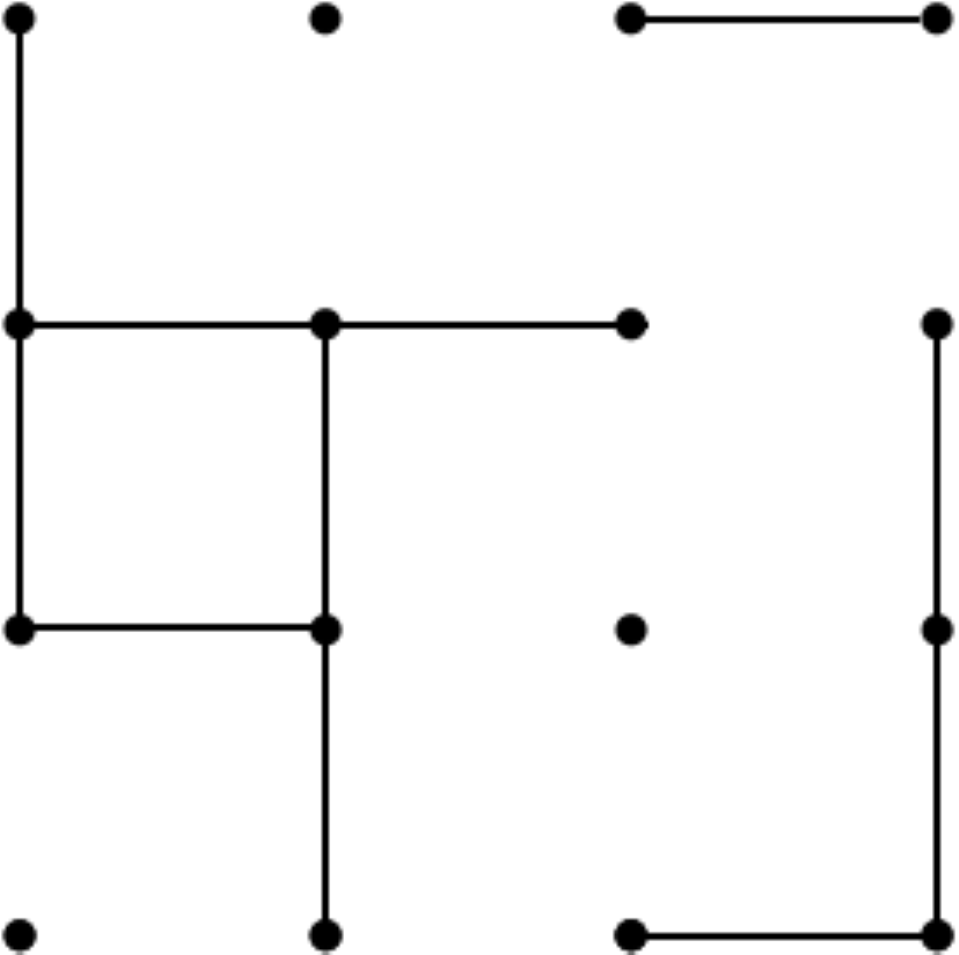} 
	\caption{An example of a random subset $X$ of bonds on a $4 \times 4$ square lattice. Here $b(X)=11$ and $c(X)=6$. Notice that isolated points count as clusters.}
	\label{fig:cluster}
\end{figure}

Let us now discuss the cluster definition of the Potts model, due to Fortuin and Kasteleyn \cite{FORTUIN1972536}. On the same lattice we consider random subsets of lattice bonds $X$. 
The probabilistic weight for a given subset $X$ to occur is defined as
\beq
\label{eq:wGamma}
w(X)=v^{b(X)} Q^{c(X)}\,,\quad v=e^\beta-1\,,
\eeq
where $b(X)$ is the total number of bonds in $X$, and $c(X)$ is the total number of \emph{clusters}---connected components in the graph which has all lattice sites as vertices and bonds from $X$ as edges. Isolated sites also count as clusters (see Fig.~\ref{fig:cluster}). The partition function is then given by: 
\beq
\label{FK}
Z_{\rm cluster}=\sum_{X} w(X).
\eeq

The factor $v^{b(X)}$ in \reef{eq:wGamma} simply means that each lattice bond is included or not into $X$ with independent probabilities $p$ and $1-p$ where $p=v/(1+v)$. 
This basic factorized probability distribution is then modified by the factor $Q^{c(X)}$.
The number of clusters $c(X)$ is a nonlocal characteristic of $X$, so definition \reef{eq:wGamma} is nonlocal. In particular, unlike for spins, it is not given in terms of a local Hamiltonian.\footnote{However some semblance of locality can be introduced by introducing the height representation, as we will review in \cite{part2}.}

For integer $Q\ge 2$ the two partition functions $Z_{\rm spin}$ and $Z_{\rm cluster}$ agree \cite{FORTUIN1972536}. To see this one perform 
a power-series expansion of $Z_{\rm spin}$ in $v$, which is the high-temperature expansion.
Using the identity
\beq
e^{\beta\delta_{s_i,s_j}}=v \delta_{s_i,s_j}+1\,,
\eeq
this expansion maps term by term onto $Z_{\rm cluster}$, with the factor $Q^{c(X)}$ arising from contracting delta-functions. In other words, for integer $Q\ge 2$ each high-temperature expansion cluster may have one of $Q$ ``colors". 

On the other hand, the cluster definition is more general as it is applicable for continuously varying $Q$. It is worth mentioning that the model is reflection positive for integer $Q$ while for non-integer $Q$ it is not (see e.g.~\cite{Biskup1998}). In this paper we will focus on real $Q>0$. 

We note in passing that the cluster definition can also be used to extract nontrivial physics from integer values of $Q=0,1$ where the spin definition would seem to be trivial. Namely, $Q = 1$ defines bond percolation, and the limit $Q \to 0^+$ corresponds to spanning trees and forests \cite{FORTUIN1972536}.

\subsection{Phase transition}
 
Consider first the integer-$Q$ Potts model defined in terms of spins. At low temperatures it has an ordered phase with $Q$ degenerate ground states where $S_Q$ is spontaneously broken and one spin value is preferred. At high temperatures we have a disordered phase where there is just one state and the spin distribution is $S_Q$-symmetric. These two phases are separated, for each $Q$, by an order-disorder phase transition at some critical temperature $\beta=\beta_c$.\footnote{The exact vaue of $\beta_c=\log(1+\sqrt{Q})$ is known in 2d from self-duality. See \cite{Wu}, including the discussion of how the duality acts in terms of the random cluster model.}
The phase transition can be detected by fixing one spin ($s_0$) and measuring the probability that a distant spin is in the same state:
\beq
p(r) = \mathbb{P}[s_r=s_0]-1/Q\,,
\eeq
where we subtract the trivial probability $1/Q$ which would arise for uncorrelated spins. An order parameter distinguishing the two phases is the value $p_\infty$ of this probability in the limit $r\to\infty$, zero in the disordered 
phase and positive in the ordered one.
 
Consider then the same phase transition within the cluster definition. At high temperatures we have $v\ll 1$ and the clusters tend to be very small. At low temperatures huge clusters occupy most of the lattice: a nonzero magnetization in the spin definition corresponds to clusters extending all across the infinite lattice in the cluster formulation. One basic cluster observable is the probability $\tilde p(r)$ that two lattice sites separated by distance $r$ lie in the same cluster, as well as an order parameter $\tilde p_\infty=\lim_{r\to\infty} \tilde p(r)$. Using the high-temperature expansion, it's not hard to show that $p(r)=(1-1/Q)\tilde p(r)$ for integer $Q$, and so the two order parameters $p_\infty$ and $\tilde p_\infty$ are equivalent (see e.g.~\cite{grimmett2006random}).

Finally let us discuss the order of the transition. We can define the correlation length $\xi$ from the rate of approach $\tilde p(r)-\tilde p_\infty\sim e^{-r/\xi}$. 
The transition is first-order or continuous depending if $\xi$ remains finite or becomes infinite at $\beta=\beta_c$.
An equivalent definition of the transition order is in terms of phase coexistence. At a first-order transition we expect that the ordered and disordered states will coexist, while at a continuous transition there is just one state.

 It can be argued using $1/Q$ expansion that the transition is first-order at $Q\gg1$ (see appendix \ref{sec:app1st}). In general we expect a critical value of $Q$ so that the transition is continuous for $Q\le Q_c$ and first-order for $Q>Q_c$. In the 2d case considered here, it is known that $Q_c=4$ \cite{Baxter:q=4}. The phase transition for $Q\le Q_c$ is described by a CFT. Much is known about this CFT as a function of $Q$, as we will see below and in \cite{part2}.
 
 Although the Potts model is usually studied on the square lattice, it is believed that the CFT describing the phase transition does not depend on the lattice type. E.g.~2d Potts models on the square and triangular lattice, as long as they have the same $Q$, should give rise to the same CFT (see e.g.~\cite{NienhuisPRL,Vernier}).\footnote{We thank Jesper Jacobsen for the following exact solvability argument which provides additional evidence. At the critical temperature, the triangular lattice Potts model can be solved mapping it onto a six-vertex Kagom\'e lattice model (\cite{Baxter}, chapter 12). Using the start-triangle relation, the latter model can be transformed to a square-lattice model (\cite{Baxter}, chapter 11). This transformation shows that correlation functions along a certain line are the same in the triangular and square Potts model, so that in particular the critical exponents are the same.} This property can be stated by saying that parameter $Q$ does not get renormalized. For integer $Q$ this can be argued by symmetry, but for non-integer $Q$ the question of symmetry appears more subtle (see the next section). 
 
 \subsection{Symmetry}
 \label{sec:sym}
As mentioned, for integer $Q\ge 2$ the Potts model has discrete global $S_Q$ symmetry. Knowing the symmetry is very useful for many reasons, for example because it allows us to identify the microscopic model with the CFT describing its fixed point. Potts models with the same symmetry will belong to the same universality class and their critical point will be described by the same CFT. 
We can change the lattice (e.g.~from square to triangular), or we can add other interactions (e.g.~next-to-nearest-neighbor). As long as symmetry is preserved, the CFT should remain the same.

Moving to non-integer $Q$, the precise mathematical meaning of symmetry is unclear. As mentioned above there is evidence that the critical point does not depend on the lattice, and this asks for a symmetry explanation. However, the group $S_Q$ certainly does not make mathematical sense for non-integer $Q$. 
Let us specify our requests for the symmetry in non-integer $Q$: it should allow a unique identification of $Q$, it should explain which perturbations of the cluster model preserve the universality class, 
and it should hopefully work for any $d$.\footnote{We note in this respect that in 2d, quantum algebra $U_{q}sl(2)$ with $q+q^{-1}=Q^{1/2}$ plays an important role in the Temperley-Lieb approach to the partition function of $Q$-state Potts model \cite{saleur1990}. This seems to be an inherently 2d phenomenon and in addition intimately related to the integrability properties of the model.} 

While we don't know of analytic continuation of $S_Q$, analytic continuation of \emph{representations} of $S_Q$
has been rigorously defined. This is called Deligne's tensor category ${\rm Rep}(S_t)$, where $t$ can be any complex number \cite{Deligne,Etingof}.\footnote{See also \cite{Deligne}, section 9, for the $O(n)$ case. We thank Alexei Borodin, Matthijs Hogervorst and Maxim Kontsevich for mentioning this work, and especially Damon Binder for extensive discussions.} 
{This construction is likely the key to formulating rigorously the symmetry of Potts model for non-integer $Q$, and we hope to return to this point in the future.} See \cite{part2} for additional speculations.

We will proceed assuming that the concept of symmetry makes sense for any $Q$, even if it's not yet been properly defined for non-integer $Q$. 

 \subsection{Weakly first-order phase transition at $Q\gtrsim 4$ and walking}
 \label{sec:2dPottsWalk}
 
Baxter \cite{Baxter:q=4} has computed the free energy of the 2d Potts model for any $Q$ at the critical temperature, 
 reducing to a 6-vertex model. From his exact solution it follows that latent heat $L$ at the transition is zero for $Q\le Q_c=4$ (continuous transition) and positive at $Q>Q_c$ (first-order transition). Baxter's solution implies that the latent heat vanishes exponentially quickly for $Q\to Q_c^+$, as
 \beq
L\sim \exp \left(-\frac{\pi^2}{2\sqrt{\delta}}\right)\,,\quad \delta=Q-Q_c\,.
\label{eq:LPotts}
\eeq
 First-order phase transitions with small latent heat (in natural units) are called weak, and the 2d Potts for $Q\gtrsim Q_c$ is an example of this.
 
Another robust characteristic of a weak first-order (WFO) transition is that the correlation length $\xi$, while remaining finite, becomes very large in the units of lattice spacing. The critical 2d Potts correlation length was computed exactly in \cite{Buffenoir}. For $Q\to Q_c^+$ the correlation length becomes exponentially large:
 \beq
 \xi\approx \xi_0 \exp \left(\frac{\pi^2}{\sqrt{\delta}}\right)\,,
\label{eq:xiPotts}
 \eeq
 with $\xi_0$ approximately constant in this limit. 

For a lattice model, to have the correlation length much larger that the lattice spacing is an example of a hierarchy, in the sense of section \ref{sec:hierarchy}. This property of the 2d Potts model at $Q\gtrsim 4$ was explained 40 years ago by Nauenberg, Scalapino, and Cardy \cite{NauenbergPRL,CNS} as a consequence of walking, in what was {perhaps} the first evocation of this mechanism in physics.\footnote{\label{note:pseudo-critical}They did not actually use the term `walking'. It seems that the mechanism does not have a standard name in statistical physics. Sometimes it is referred to as `pseudo-critical behavior' \cite{Nahum:2015jya,Wang:2017txt,Sandvik}.}

Let us review this explanation and the evidence in its favor. The key assumption is that the RG evolution is described by Eq.~\reef{eq:RGwalk}, with parameter $y$ a monotonic function of $Q$. $Q=Q_c$ corresponds then to $y=0$, and one assumes that near this point $y$ has approximately linear dependence on $\delta$:
\beq
y=C\, \delta +O(\delta^2)\,.
\label{eq:ydelta}
\eeq
The constant $C$ must be positive, so that $y>0$ (no fixed point at real $\lambda$) corresponds to $Q>Q_c$. The value of $C$ can be readily fixed by demanding that the hierarchy \reef{eq:xiWalk} reproduce the exactly known correlation length asymptotics \reef{eq:xiPotts}. One gets\footnote{This step was not done in  \cite{NauenbergPRL,CNS} because the correlation length asymptotics was not known at the time. They arrived at the same value of $C$ via the exactly known energy operator dimensions and the latent heat asymptotics, see below.}
\beq
C=1/\pi^2\,.
\label{eq:cex}
\eeq

Consider then what happens for $Q<Q_c$. It is convenient to enlarge the coupling space of the Potts model by considering the \emph{dilute Potts model}. In this model the Potts spins or clusters live only on a part of the lattice, while the rest is occupied by vacancies.\footnote{This is also called the Blume-Emery-Giffiths model \cite{Blume,Berker}. For a cluster definition applicable at non-integer $Q$ see \cite{FK_vacancies}.} One can also think that vacancies are generated by RG transformations and represent disordered spin configurations \cite{NienhuisPRL}.

Now, it is known that the dilute Potts model has for $Q<Q_c$ two fixed points. One of them is the same as the critical point of the usual non-dilute Potts model. The other fixed point is tricritical, obtained by tuning both the temperature and the chemical potential for the vacancies. Ref.~\cite{NienhuisPRL} first found these fixed points by means of an approximate RG transformation, and showed that they annihilate at $Q=Q_c$. This picture of two fixed points at $y<0$ annihilating at $y=0$ agrees with section \ref{sec:walking} (see Fig.~\ref{fig:QCDvsPotts}(b) below). We therefore identify the more stable $\lambda=\lambda_-$ fixed point as the critical and $\lambda=\lambda_+$ as the tricritical Potts model.

In fact, much is known about the CFTs describing these fixed points, and this can be used to further check and complete the walking RG picture.\footnote{The operator spectrum of both CFTs is fully known, some of it via exact lattice solution \`a la Baxter, and the rest via Coulomb gas \cite{diFrancesco:1987qf}. The OPE coefficients are known fully for $Q=2,3,4$ and partially for other $Q$'s.
	 We will review and use this information in \cite{part2}.}  Here we will use the two lowest singlet operators $\vareps$ and $\vareps'$, referred to as the leading and subleading temperature perturbations.
Their dimensions are known exactly for $Q<Q_c$ \cite{denNijs,Black-Emery,Nienhuis1982a}, with the following asymptotics at $Q\to Q_c^-$:
\begin{gather}
\Delta_\vareps=\frac 12\mp \frac 3{4\pi}\sqrt{|\delta|}+O(\delta)\,,\label{eq:vareps}\\
\Delta_{\vareps'}=2\mp \frac 2{\pi}\sqrt{|\delta|}+O(\delta)\,.\label{eq:vareps'}
\end{gather}
(The upper sign corresponds to the tricritical fixed point.) Operator $\vareps'$ is close to marginality and should be identified with the operator $\calO_\lambda$, whose dimension is predicted by RG in \reef{eq:Dpm}. Using \reef{eq:ydelta} and \reef{eq:cex}, we see that the RG prediction \reef{eq:Dpm} agrees with the exact result \reef{eq:vareps'}.

Operator $\vareps$ is strongly relevant, and its coupling (denoted $\phi$ in \cite{NauenbergPRL,CNS}) must be finetuned to zero to reach the phase transition. This is a particularity of the Potts model compared to the basic scenario in section \ref{sec:walking} and to walking in gauge theories.
Small deviations of $\phi$ from zero are governed by an RG equation of the leading form
\beq
d\phi/dt = -(a+ b\,\lambda)\phi+\ldots\,.
\label{eq:RGphi}
\eeq
The fixed point dimension of $\vareps$ is then given by
\beq
\Delta_\vareps=2-(a+b\,\lambda_\pm) +O(\delta)\,.
\eeq
Demanding agreement with \reef{eq:vareps} allows us to fix the two constants $a$, $b$:
\beq
a=\frac 3 2,\quad b = \frac3 4\,.\label{eq:ab}
\eeq

The RG equation \reef{eq:RGphi} is also important when studying the WFO regime $y>0$. The running of $\phi$ must be taken into account when computing the latent heat (which is the derivative of free energy w.r.t.~$\phi$). The computation of \cite{NauenbergPRL,CNS} finds an exponentially small latent heat of the same form as \reef{eq:LPotts}. Precise agreement in the exponent is obtained for $a=3/2$ as in \reef{eq:ab}, providing yet another consistency check of this picture.

\subsection{Lessons and questions}
\label{sec:lessons}
We see that the 2d Potts model presents a remarkable opportunity for testing the idea of walking.
Not only some aspects of it are exactly solvable, it's also relatively easy to study via Monte Carlo simulations. The key assumption is that the same RG equations \reef{eq:RGwalk}, \reef{eq:RGphi} apply on both sides of $Q=Q_c$ provided that we make the parameter $y$ depend on $Q-Q_c$ as in \reef{eq:ydelta}. Coefficients in these equations can be fixed demanding the consistency with the exactly known critical exponents at $Q<Q_c$. Solving the same equation for $Q>Q_c$, one obtains approximate results for the correlation length and latent heat in the phase where the transition is weakly first-order, which can be checked against the exact solution on the lattice.

\begin{table}\centering
	\begin{tabular}{rrrrrrr}\toprule
		$Q$ & 5 & 6 & 7 & 8 & 9 &  10 \\
		$\xi$ & 2512.2 & 158.9 & 48.1 & 23.9 & 14.9 & 10.6 \\
		\bottomrule
	\end{tabular}
	\caption{The 2d Potts model correlation length on the square lattice at the first-order phase transition for $Q=5$ - 10, computed from \cite{Buffenoir}, Eq.~(4.46).
	}
	\label{tab:xiPotts}
\end{table}

 {What is the range of $Q$ for which walking behavior persists?
Looking at Table \ref{tab:xiPotts}, we see large correlation lengths up to $Q\lesssim 10$. Eq.~\reef{eq:xiPotts} works pretty well in this whole range, provided that one allows the coefficient $\xi_0$ to vary by 30\%, from 0.13 to 0.19 (while $\xi$ itself varies by factor 250). Naively this is puzzling, as it may seem that the expansion in $Q-Q_c$ has an unexpectedly large range of validity. However, as mentioned in section \ref{sec:walking}, the true criterion for walking should involve nonperturbative information about complex CFTs, rather than $Q-Q_c$.
See section \ref{ssec:complexCFTS_walking} for a general discussion and \cite{part2} for details specific to the 2d Potts.}

\section{Walking in high energy physics: 4d gauge theories}
\label{sec:BZ}

Slowly running or walking coupling was first introduced in particle physics in the context of technicolor models of electroweak symmetry breaking \cite{Holdom:1981rm,Yamawaki:1985zg,Appelquist:1986an}. These models later received the name ``walking technicolor" (WTC). Here we focus on the simplest setup where walking is supposed to occur. Connection to the electroweak phenomenology will be commented upon in section \ref{sec:pheno} below. 

This simple setup is the 4d gauge theory with gauge group $SU(N_c)$  and $N_f$ massless fermions in the fundamental representation. We will denote $x=N_f/N_c$ and assume $x<x_{\rm AF}=11/2$ so that the theory is asymptotically free. It is believed that in an interval of $x$ below $x_{\rm AF}$, 
\beq
x_{c}<x<x_{\rm AF}\,, \label{BZ}
\eeq
this theory flows at long distances to a CFT called the Banks-Zaks (BZ) fixed point \cite{Belavin:1974gu,Caswell:1974gg,Banks:1981nn}. The interval \reef{BZ} is called the conformal window.

The BZ fixed point is weakly coupled near $x_{\rm AF}$ but is strongly interacting near $x_c$. 
To consider the weakly coupled $x\to x_{\rm AF}$ limit we can formally consider $N_f$ to be a continuously varying parameter. It's not clear if this makes full sense non-perturbatively.\footnote{On the other hand, the Potts model with $Q$ states discussed above can be defined non-perturbatively for continuously varying $Q$, although it's unitary only for integer $Q$.} If we wanted to be conservative, we could restrict $N_f$ to be an integer, but then we would have to take large $N_c$. For simplicity, we will not keep track of large $N_c$ counting. In any case, it is believed that the conformal window is non-empty also for finite $N_c$. There is evidence for that from various theoretical considerations and from lattice Monte Carlo simulations, see 
e.g.~\cite{Nogradi:2016qek} for a review.\footnote{Here we are focusing on 4d gauge theories, but a similar story is believed to hold in 3d gauge theories coupled to $N_f$ massless fermions. The difference is that in this case conformal window exists both for the abelian and nonabelian gauge theories, and that it extends all way to $N_f=\infty$. The lower boundary of the conformal window is not known in 3d just like in 4d. A natural possibility is that the 3d conformal window terminates via fixed point annihilation, as discussed below for the 4d case. See \cite{Giombi:2015haa,Gukov:2016tnp} for recent discussions and references to prior work. The present-day lattice QCD community is neglecting the 3d case, with a few notable exceptions such as \cite{Karthik:2016ppr,Karthik:2018nzf}. We find this neglect regrettable and methodologically wrong. Not only is the 3d conformal window interesting in its own right and has multiple connections to contemporary condensed-matter physics, see section \ref{sec:further}, it is certainly easier than the 4d case, and should be solved first as a warmup.}

\begin{figure}
	\centering
	\begin{subfigure}{.5\textwidth}
		\centering
		\includegraphics[width=.9\linewidth]{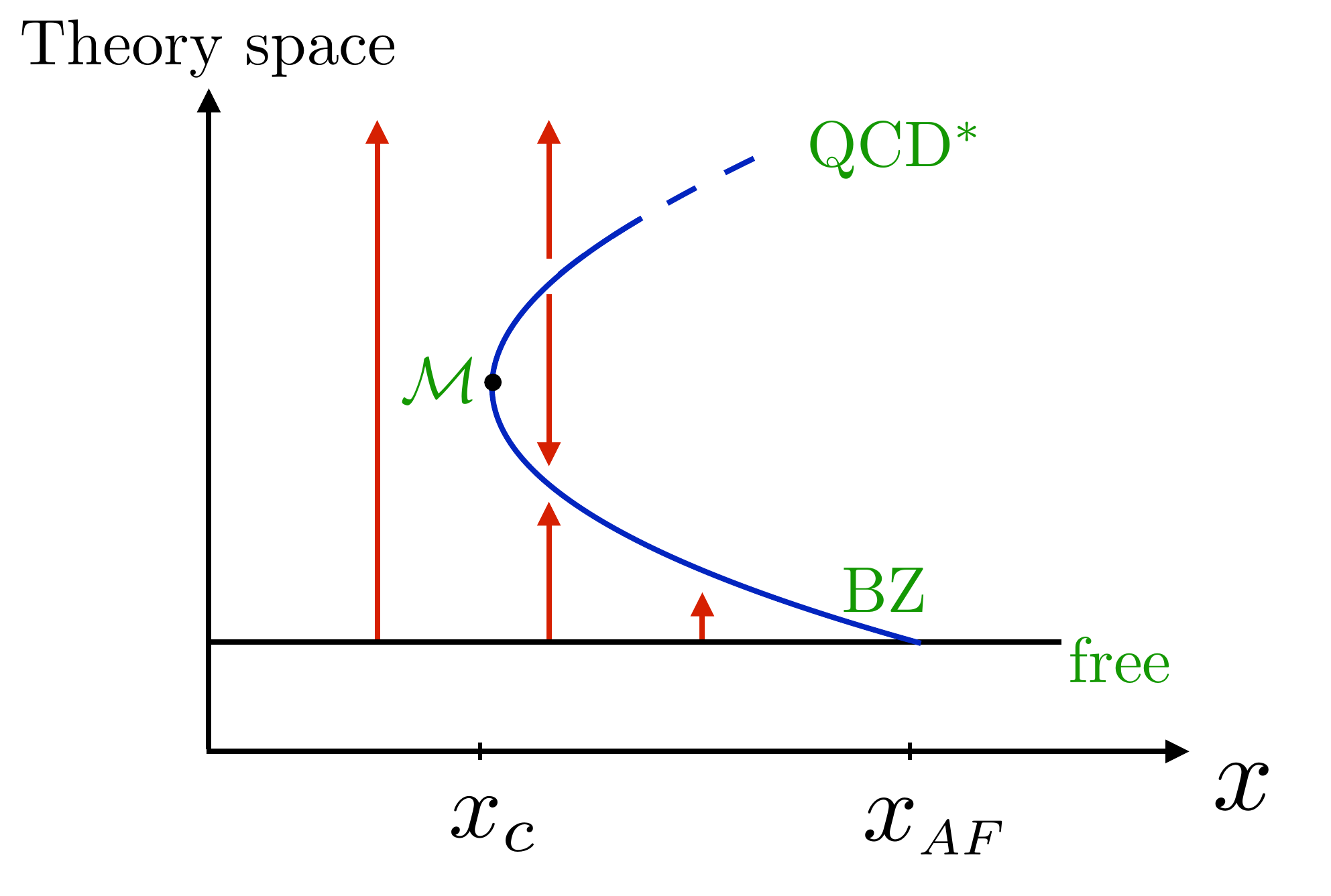}
			\caption{}
	\end{subfigure}%
	\begin{subfigure}{.5\textwidth}
		\centering
		\includegraphics[width=.9\linewidth]{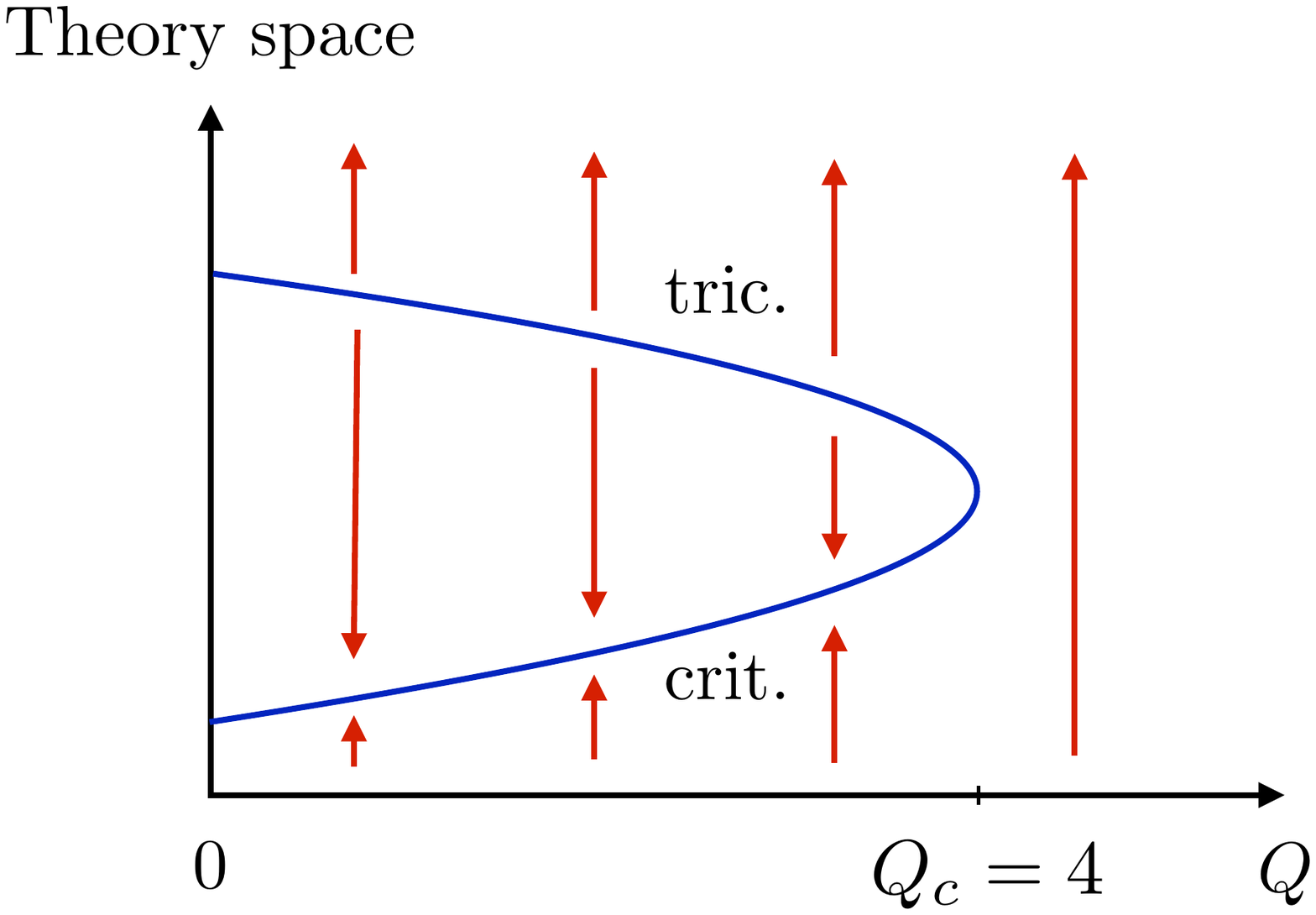}
		\caption{}
	\end{subfigure}
	\caption{{\it Left:} schematic view of the space of existing fixed points of 4d gauge theories as a function of $x=N_f/N_c$. The trivial fixed point (free) exists for 
		any $x$. The line of BZ fixed points branches off from the free theory line at $x=x\AF$. At some smaller $x=x_c$ it annihilates with the line of QCD$^*$ fixed points. This latter line should exist for $x$ close to $x_c$ but it's not a priori clear where it starts. {\it Right:} schematic view of the space of fixed points for the 2d Potts model. No theory merges with the free theory, at least in the range $Q>0$ we are interested in.
	}
	\label{fig:QCDvsPotts}
\end{figure}

On the other hand, for $x<x_c$ there is no fixed point, the theory instead flowing to a confining phase with spontaneously broken chiral symmetry. So the fixed point must disappear as $x$ approaches $x_c$ from above. One possibility is that the fixed point annihilates with another fixed point; see Fig.~\ref{fig:QCDvsPotts}(a). This is the scenario advocated in \cite{Kaplan:2009kr}, where the new fixed point is called QCD$^*$ (see also \cite{Gies:2005as}). One necessary condition for this scenario is that QCD$^*$ have the same global symmetry $SU(N_f)\times SU(N_f)\times U(1)$ as the BZ fixed point with which it is annihilating. 

Fixed point annihilation appears to us the simplest scenario which explains fixed point disappearance and confining behavior below $x_c$. In what follows we assume that this scenario is realized. We will comment on other logical possibilities in section \ref{sec:other-ends} below.

We call $\calM$ the common endpoint of the BZ and QCD${}^*$ branches:
\beq
\calM=\text{BZ}(x_c)=\text{QCD}^*(x_c)\,.
\eeq
Let us look more carefully at how the annihilation happens. For $x$ slightly larger than $x_c$ there is an RG flow that connects BZ and QCD$^*$. Close to $\calM$ this RG flow is very short and consequently can be described within conformal perturbation theory around QCD$^*$. Call $\calO$ the operator that induces this flow, and $\lambda$ the corresponding coupling constant. Since the flow degenerates when we approach $\calM$,
$\calO$ should become marginal at $x=x_c$:
\beq
\label{eq:firstpred}
\Delta_\calO=d\quad(x=x_c)\,. 
\eeq
It is the first robust prediction of this scenario. 

Next, let's discuss the beta-function. The request of two fixed points for $x>x_c$ and no fixed points for $x<x_c$ naturally leads to the beta-function of the by now familiar form \reef{eq:RGwalk}, where $y\approx C \delta$ for small $\delta = x-x_c$. This can be compared to Eq.~\reef{eq:ydelta} for the Potts model where $\delta=Q-Q_c$. However, the precise value of $C$ is not known for the 4d gauge theory case, only its sign is fixed: $C<0$, to have a conformal phase for $x>x_c$ (the opposite of that of the Potts model).

At this point much of the discussion of section \ref{sec:walking} carries over. For example for $x$ slightly above $x_c$ the dimension of $\calO$ should have characteristic square root behavior \reef{eq:Dpm}.
This square root behavior was emphasized recently in 
\cite{Gukov:2016tnp}. 
On the other hand, for $x$ slightly below $x_c$, we will have walking behavior resulting in the hierarchy \reef{eq:xiWalk}.\footnote{This scaling in the context of gauge theories is also known as ``Miransky scaling" \cite{Miransky1985}.} \footnote{We thank Igor Klebanov for pointing out that beta-function of the walking form \reef{eq:RGwalk} was also obtained in \cite{Dymarsky:2005uh} for orbifolds of $\mathcal{N}=4$ Super Yang-Mills (SYM) theory. In their case the parameter $y=-D\lambda$, where $\lambda$ is 't Hooft coupling and $D$ is a combination of dynamical characteristics of the theory. If $D<0$, one can reach the walking regime by dialing $\lambda$ small. The existence of a dialable coupling $\lambda$ makes this example somewhat similar to the BKT transition discussed in appendix \ref{sec:BKT}. It should be contrasted with the theories discussed in this paper where a symmetry parameter is varied to reach the walking regime.}

Let us point out some important differences between the scenario we are describing and the descriptions of walking that have appeared previously in the particle physics context. First, unlike in the first references on WTC, we see that it would be misleading to think of the slow coupling $\lambda$ as the gauge coupling constant $g$. 
 The gauge coupling controls the flow next to the free theory. Near $\calM$, the flow is strongly coupled, and $\lambda$ is one particular linear combination of infinitely many couplings which happens to control the fixed point annihilation. Here we agree with Ref.~\cite{Kaplan:2009kr} which did not postulate relation between $\lambda$ and the gauge coupling. Further insight into the distinction between $g$ and $\lambda$ can be gained by considering theories with large-$N$ counting, see appendix \ref{sec:largeN}. As we explain there, in this case $\calO$ has to be a double-trace operator, so it cannot even mix with the operator controlling the gauge coupling, $\tr FF$, at the leading order.

On the other hand, our discussion differs from Ref.~\cite{Kaplan:2009kr} in that they specialized early on to the large $N_c$ limit, as well as used the holographic analogy. An important role in their discussion is played by an operator of dimensionality $d/2$ present at the fixed point $\calM$ in addition to the marginal operator $\calO$. In the old WTC literature, it also often stated that the $\bar \psi \psi$ operator has scaling dimension $d/2=2$ at the onset of the walking regime (see e.g.~\cite{Cohen:1988sq}). Our discussion on the other hand tried to separate the general phenomenon of walking from its large $N$ limit which may have special features. The operator of dimensionality $d/2$ is one of such features, as we discuss in appendix \ref{sec:largeN}. We do not find any support for the existence of such an operator in a generic case. In fact in the Potts model in two dimensions no operator of dimension one is present in the spectrum at $Q=4$ when two fixed points annihilate, see \cite{part2}.

We would also like to stress that
walking is not a finetuned scenario. First, as we argued in section~\ref{sec:NatWalk}, walking does not involve any finetuning price whatsoever as far as the initial conditions of RG flow are concerned. One may be worried about finetuning in theory space, i.e.~how close $x$ should be taken to $x_c$. By analogy with $Q-4$ for the 2d Potts model, we believe that walking may occur for $x$ deviating from $x_c$ by $O(1)$. In particular, in our opinion there seems to be no reason preventing walking to occur for some integer $N_f$ even if $N_c$ is small.

Finally, we would like to contrast the hierarchy appearing due to walking in massless QCD at $x\lesssim x_c$ with a different sort of hierarchy which would appear in the BZ regime \reef{BZ} if we perturbed the massless QCD with a fermion mass term. Suppose we add a very small equal mass for all fermions. The resulting theory would first go close to a BZ fixed point, sit near it while the fermion mass term keeps growing, and finally flow out to a gapped phase. This hierarchy generation is classified as the tuning scenario of section \ref{sec:Tuning}; it is technically natural tuning because the mass term breaks chiral symmetry. We included it here to emphasize that this is not what walking is about.

\subsection{Walking and the electroweak phenomenology beyond the Standard Model}
\label{sec:pheno}

We would like to make here a few comments concerning connections between walking and electroweak phenomenology. As is well known, WTC was originally hypothesized in the hope of making less severe the problem of flavor changing neutral currents present in technicolor models of electroweak symmetry breaking.
It is not our goal here to review this story in detail. For a review in the context of RG flows we refer to \cite{Luty:2004ye}. Relevantly for us, WTC models contain a sector which exhibits the walking regime of RG flow in a range of energies above the electroweak scale. This allows, in this range of energies, to classify operators belonging to the ``walking" sector according to their (approximate) scaling dimensions. In particular, this sector is assumed to contain an operator with the quantum numbers of the Higgs field (in explicit models it is a fermion bilinear operator $\bar \psi\psi$). Various phenomenological constraints then require that this operator has dimension close to one and at the same time does not behave as a free field. 

A proposal made in \cite{Luty:2004ye} was that we can treat the walking theory as a strongly interacting unitary CFT perturbed by a weakly relevant operator resulting in a slow RG flow, called ``conformal technicolor" scenario. Subsequently, Ref.~\cite{Rattazzi:2008pe} pointed out that unitary CFTs with restrictions on operator dimensions required to make conformal technicolor scenario phenomenologically viable can be constrained from first principles using the conformal bootstrap techniques. The resulting constraints were fully worked out in \cite{Poland:2011ey}, with rather pessimistic conclusions: the generic scenario was found inconsistent.\footnote{However, consistency could be saved by making special assumptions about the structure of the flavor sector on top of admitting some tuning.} 

There is however a loophole. Conformal technicolor scenario of \cite{Luty:2004ye} is a meaningful scenario, but it should not be considered as an equivalent formulation of walking. Indeed, according to our classification in section \ref{sec:Tuning} it belongs to ``mild tuning". Walking is radically different in that there are no unitary fixed points. For this reason, strictly speaking, conformal bootstrap results derived in \cite{Rattazzi:2008pe} and \cite{Poland:2011ey} under the assumption of unitarity cannot be used to rigorously bound the approximate scaling dimensions in the walking regime. 

 One may be wondering how serious this loophole actually is. After all, even in the walking scenario the flow passes near a CFT, which while not being exactly unitary can be described as ``almost unitary" since imaginary parts of operator dimensions have to be small for the walking behavior to take place, Eq.~\reef{eq:delta_y}. Can bootstrap bounds get significantly relaxed in presence of such small violations of unitarity? There is no immediate answer to this question, and this problem should be considered open. One related example where bounds seem to get relaxed will be discussed in section \ref{sec:further}.

In retrospect, it's fortunate that the loophole was not noticed at the time of writing \cite{Rattazzi:2008pe},\footnote{One of us (S.R.) only realized it in 2016.} since otherwise the conformal bootstrap methods may not have been developed, and those methods since then found many other applications \cite{review}.

\subsection{Light dilaton?}
\label{sec:dilaton}
Finally, let us mention that since the early days the presence of a parametrically light ``pseudo-dilaton" is believed to be present in WTC models \cite{Yamawaki:1985zg}. This is supposed to be a resonance with quantum numbers $0^{++}$, and its parametric lightness means that it's much lighter than the other massive technihadrons (e.g.~the technirho), the mass ratio going to zero as one approaches $x_c$ from below. This belief is a popular subject of the lattice studies, see e.g.~\cite{DeGrand:2015zxa}. We therefore find it necessary to comment on this issue. 

In our opinion, walking behavior is not by itself a sufficient condition for the presence of the pseudo-dilaton, that is a pseudo-Goldstone boson of spontaneously broken scale invariance, but it requires further assumptions.\footnote{See \cite{Coradeschi:2013gda} for the only known to us field-theoretical construction of a naturally light pseudo-dilation, which involves a host of dynamical assumptions very different from the situation at hand.} The presence of a Goldstone boson requires {\it spontaneous} breaking of a symmetry, while walking behavior corresponds to a small, in some sense, but {\it{explicit}} breaking.\footnote{This is especially apparent from our description of walking as a perturbation of a (complex) CFT by an operator, which will be introduced in section \ref{ssec:complexCFTS_walking}. Addition of this operator to the action explicitly breaks the symmetry.} It is the possibility of spontaneous breaking of conformal symmetry which represents an extra assumption. To have this possibility, we need a CFT with a moduli space of vacua, which is a phenomenon logically completely independent of walking. 

Consider for example the CFT $\calM$ describing the annihilation point of the BZ and QCD${}^*$ branches. This is certainly a special theory, as evidenced by the fact that it contains a marginal operator $\calO$.
If it had a moduli space, there would be room for realizing a light pseudo-dilaton scenario at $x<x_c$. However, does $\calM$ have a moduli space of vacua? We do not actually believe that this is the case. 

Let us recall what is known about CFTs with a moduli space.
In known examples the moduli space is parametrized by giving expectation values to scalar fields.
The simplest example is the free massless scalar $\phi$ in $d>2$ dimensions, where we can give an arbitrary expectation value to $\phi$.\footnote{Notice that this theory in 4d also contains a marginal operator $\phi^4$. But it would of course be a logical fallacy to conclude that any 4d theory with a marginal operator, like $\calM$, should have a moduli space. There are many examples of supersymmetric CFTs which have a conformal manifold (i.e.~exactly marginal deformations), but do not have a moduli space of vacua. A very simple example is discussed in \cite{Baggio:2017mas}. } Passing to the interacting case, all known interacting CFTs with a finite number of degrees of freedom having a moduli space are supersymmetric. One example is the $\calN=1$ supersymmetric QCD which will be discussed in section \ref{sec:other-ends} below. The key to the existence of the moduli space of supersymmetric theories is the nonrenormalization theorem for the superpotential. For non-SUSY theories we generically expect that flat directions, even if present in the UV Lagrangian, are lifted by quantum effect so that the IR fixed point has no moduli space.

Constructions based on holography often lead to large-$N$ `CFTs' which appear to possess a moduli space of vacua, usually parametrized by a radial location of a brane. Notice however that only in supersymmetric cases (like the $\calN=4$ super Yang Mills) do these constructions correspond to UV complete theories. In non-SUSY cases these `CFTs' are at best effective theories and it's usually not known if they can be UV completed. Non-supersymmetric small-$N$ CFTs are not expected to have weakly coupled gravitational duals. 

A related example of how moduli space can be a large $N$ artefact is as follows. A non-holographic non-SUSY model having a moduli space of vacua in the strict $N=\infty$ limit was discussed in \cite{Bardeen:1983rv}. However, it was shown \cite{David:1985zz,Omid:2016jve} that this moduli space does not survive at finite $N$.

To summarize, we don't see any theoretical evidence for the presence of the light pseudo-dilation in the walking regime of non-SUSY gauge theories. However, we do find the lattice investigations of gauge theories near the end of the conformal window intriguing, notwithstanding the issue of the light dilaton. It is a hard subject and should be done with care \cite{Nogradi:2016qek}.

For completeness let us discuss what happens for the 2d $Q>4$ Potts model. In this case the spectrum of massive excitations close to the end of `conformal window', i.e.~for $Q=4+\eps$, is known exactly from integrability \cite{Delfino:2000xt}. It consists solely of kink-like excitations with masses of order of the infrared strong coupling scale. Of course when $\eps\to0^+$ these excitations become massless, but this happens for all the IR excitations in the model simultaneously, so that the mass-ratios stay finite. We believe that this is what should happen near the end of the conformal window in QCD as well (except for the goldstone bosons which should stay exactly massless of course). We have not seen any evidence to the contrary.\footnote{Recent Ref.~\cite{Appelquist:2018yqe} studied $SU(3)$ gauge theory with $N_f=8$ massless fermions in the fundamental via lattice Monte Carlo. They assign it close to, but somewhat below, the end of the conformal window and look for signs of light pseudodilaton. In the volumes they study, pseudoscalars $\pi$ as well as the scalar $0^{++}$ have mass $\approx 0.5$ (in the units in which $m_\rho=1$). We take it as a sign that their simulations are still far from the infinite-volume limit in which $m_\pi/m_\rho$ should go to zero. We conservatively predict that as one goes to larger volumes $m_{0^{++}}/m_\rho$ will plateau, and there will be no light pseudodilaton. 
	We thank Anna Hasenfratz for discussions.}

It should be said that in 2d the general belief is that moduli space of vacua is impossible even for supersymmetric CFTs, due to infrared effects not unlike those which prevent the spontaneous breaking of global continuous symmetry in 2d via the Coleman-Mermin-Wagner theorem.\footnote{In the literature on 2d SUSY CFTs one uses the term `moduli space' in a different meaning, to denote the conformal manifold, i.e.~the space of all exactly marginal deformations.} This goes towards showing once more that walking and spontaneous breaking of conformal invariance are logically independent phenomena.

\subsection{Other possibilities for the end of conformal window}
\label{sec:other-ends}

The above discussion was based on the simplest assumption \cite{Kaplan:2009kr} that the BZ line of fixed points terminates by annihilating with another line called QCD${}^*$ and then moves to the complex plane. 
Let us now discuss what are possible other ways for the transition from the BZ regime at $x$ near $x_{\rm AF}$ to the chiral symmetry breaking ($\chi$SB) regime for $x<x_c$.

One could imagine a possibility that the BZ fixed point, which appears by splitting off the free theory, disappears by the inverse of this process, i.e.~by merging with another free theory. As we will see below this is likely inconsistent with the $\chi$SB phase for $x<x_c$, but it's instructive to discuss this anyway. We can imagine the RG flow happening in the space of all theories having $G=SU(N_f)\times SU(N_f)\times U(1)$ global symmetry. In this theory space there are special points: free gauge theories. The beta-functions vanish at free theory points, and we may imagine that they are rather generic vector fields in the bulk of the theory space. Under this genericity assumptions, the fixed points can naturally appear or disappear through two processes: split off or merge with a free theory,\footnote{Ref.~\cite{Kaplan:2009kr} also considered a possibility which they call ``running off to infinite coupling". We prefer to use a terminology which is invariant under reparametrizations of the coupling space. What matters is not whether the coupling is finite or infinite, but whether the point where the topology of the RG flow changes is a truly special point of the coupling space, e.g.~if it corresponds to a free theory in terms of some dual variables, as it happens for the $\calN=1$ SQCD discussed below.} or pair create and annihilate in the bulk of the theory space. (See below for a third process involving global symmetry enhancement.)
The genericity argument can be made mathematically precise in the context of bifurcation theory for finite-dimensional families of vector fields \cite{Gukov:2016tnp}. 

\begin{figure}
	\centering
	\includegraphics[width=.7\linewidth]{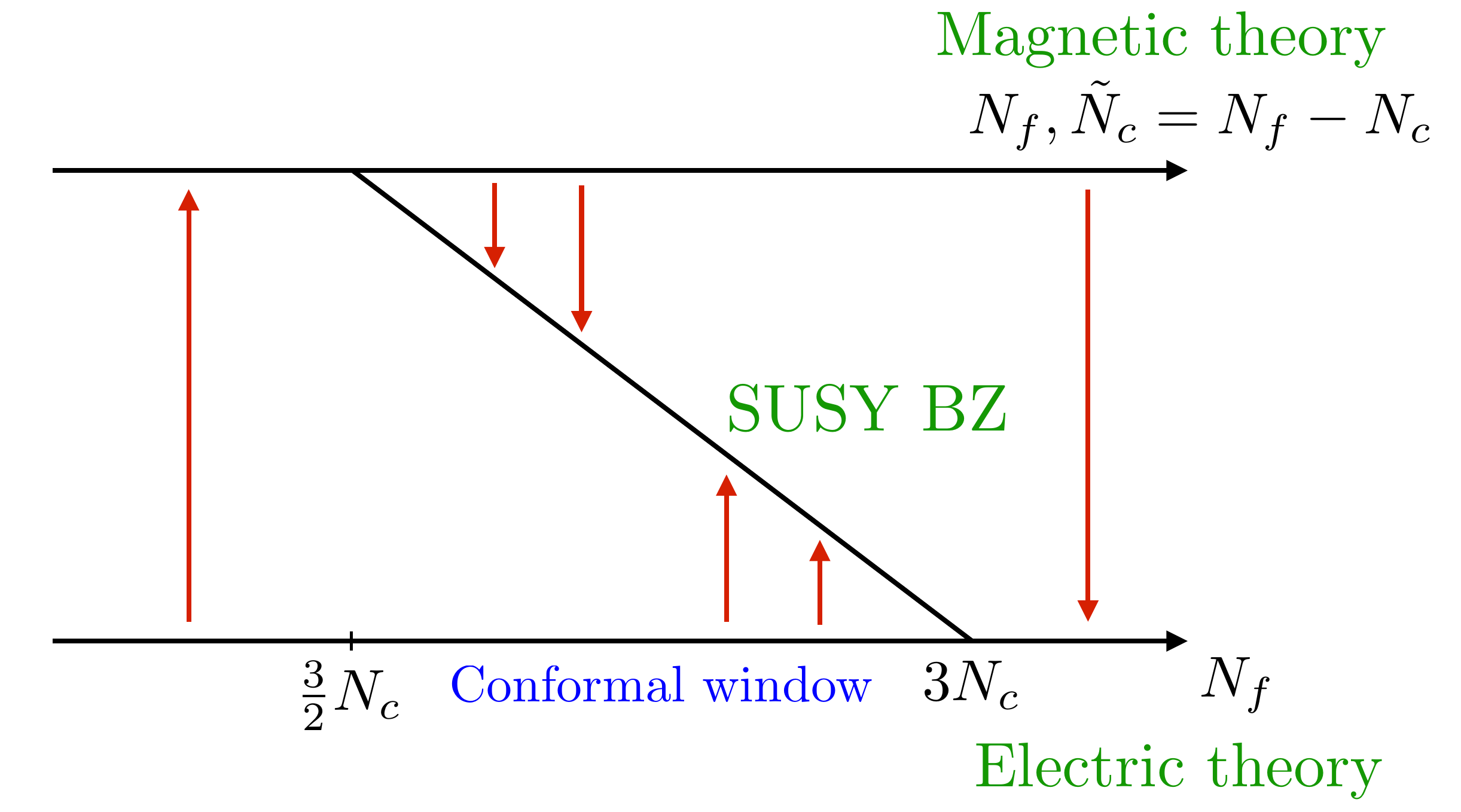}
	\caption{Conformal window for the $\calN=1$ SUSY case. In this case the BZ-like fixed point disappears by merging with another free theory. Compare with Fig.~\ref{fig:QCDvsPotts} in the non-SUSY case.}
	\label{fig:scenario}
\end{figure}

Disappearance via merging is realized in $\calN=1$ supersymmetric QCD (see Fig.~\ref{fig:scenario}). The extent of the conformal window in this theory is exactly known: $3/2<x<3$.\footnote{In connection with the discussion in section \ref{sec:dilaton}, we note that the SUSY CFT describing the IR fixed point allows spontaneous breaking of conformal invariance, as it has a moduli space of vacua of complex dimension $2N_f N_c-(N_c^2-1)$.} The BZ-like fixed point disappears at $x=3/2$ via merging with another free gauge theory, with a different number of colors $N'_c=N_f-N_c$, a manifestation of Seiberg duality \cite{Seiberg:1994pq}. 
 In the SUSY case for $x$ just below $3/2$ there is no chiral symmetry breaking, the theory instead flowing to a free magnetic phase. {On the other hand in the non-SUSY case we expect chiral symmetry breaking below $x_c$. 
Therefore the non-SUSY BZ fixed point cannot disappear via merging with free theory, and SUSY intuition is not a good guidance for this particular question.}\footnote{Ref.~\cite{Ryttov:2017lkz} studied the conformal window of non-SUSY gauge theories in perturbation theory around $x_{\rm AF}$ and used the criterion $\Delta_{\bar \psi\psi}=1$ (unitarity bound) for the lower end of the conformal window. This was inspired by the SUSY case, where $\bar \psi\psi$ becomes free scalar at the lower end as a consequence of merger with free theory. Since as we described in the non-SUSY case there is no merger with free theory, their criterion is inadequate in the non-SUSY case. Moreover, whatever the criterion, we believe perturbation theory is inadequate to describe the lower end of conformal window where anomalous dimensions become $O(1)$. We thank Robert Shrock for an email about his work.}

On the other hand one could ask if the QCD${}^*$ line could merge with a free theory at some $x_*>x_c$, which may or may not be equal to $x_{\rm AF}$. If so we could get QCD${}^*$ as an RG fixed point flowing from that free theory, which would be weakly coupled for $x$ near $x_*$. This problem is constrained by the requirements that QCD${}^*$ should have the same symmetry as BZ, that it should have exactly one relevant singlet scalar operator, and also by 't Hooft anomaly matching. Ref.~\cite{Kaplan:2009kr}
tried a few RG flows but they either did not manifestly have the requisite symmetry, or did not yield a fixed point. So this problem is open.

The final possibility that we would like to mention involves global symmetry enhancement. Namely, imagine that the BZ fixed point, when moving around in the theory space as a function of $x$ collides for $x=x'$ with a fixed point BZ$'$ which is interacting but which has a strictly larger global symmetry $G'\supset G$.
In such a situation we can have an exchange of stability, i.e.~the BZ fixed point is stable for $x>x'$, while BZ$'$ is stable for $x<x'$. This can happen naturally at a point of the BZ$'$ line where some operator which breaks $G'$ to $G$ crosses marginality. 
A well-known example of this phenomenon is the collision between the cubic and the $O(N)$ fixed points of multifield scalar theories in $4-\eps$ dimensions, which pass through each other interchanging stability at some $N_c=4+O(\eps)$ (\cite{Aharony} and \cite{Pelissetto:2000ek}, section 11.3).

Notice that the collision with free theories discussed above can also be viewed as an example of symmetry enhancement, since free theories possess higher spin symmetries. In this case, in perturbation theory the fixed point `goes through', but the fixed point coupling on the other side has a bad sign. E.g.~if we tried to formally continue BZ fixed point to $x>x_{\rm AF}$ we would find a theory at negative squared gauge coupling $g_*^2$. In a related case of trying to continue the Wilson-Fisher fixed point of $\lambda\phi^4$ to $4+\eps$ dimensions we would find a theory with negative $\lambda_*$. The common lore is that these fixed points are not well defined nonperturbatively. Indeed from the path integral point of view it's hard to believe that they make sense. So in this case we speak of fixed point merger as opposed to intersection. 

\begin{figure}
	\centering
	\includegraphics[width=.7\linewidth]{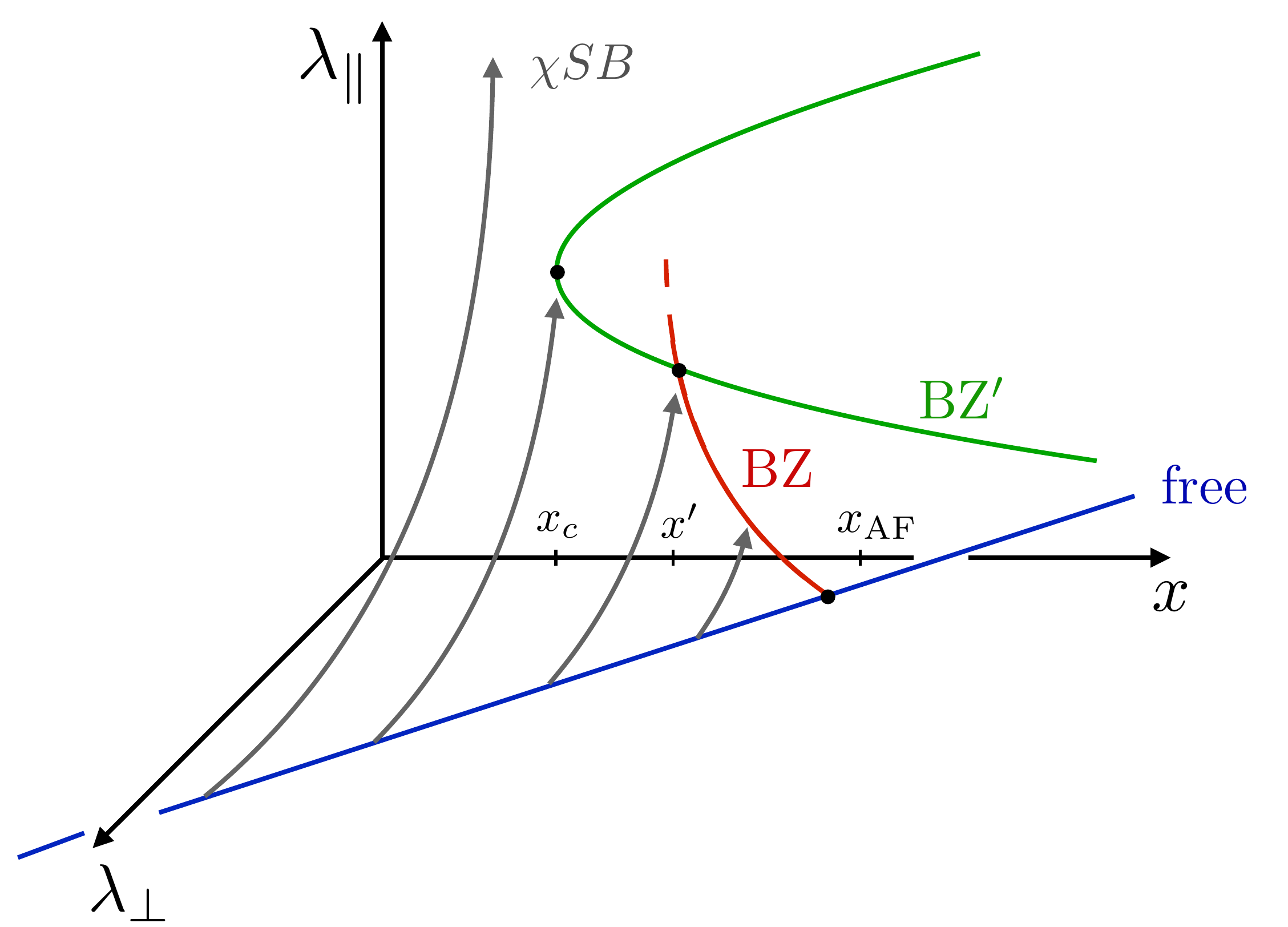}
	\caption{The non-minimal scenario. We call collectively by $\lambda_\|$ the couplings which preserved $G'$ symmetry, and by $\lambda_\perp$ the couplings which break $G'$ to $G$. The line of fixed points BZ${}'$ lives at $\lambda_\perp=0$. The RG flow lines lie in planes of constant $x$.}
	\label{fig:scenario1}
\end{figure}

On the contrary, if BZ collides with BZ$'$, we may indeed expect an intersection, because an interacting BZ$'$ is not expected to be as fragile as free theories, and the line of BZ fixed points should make sense on its both sides.\footnote{As a technical comment, we note that the actual fate of the BZ fixed point in this scenario at $x<x'$ depends on the $G'$ quantum numbers of the operator $\calO$ which breaks $G'$ to $G$ and induces a flow between BZ and BZ${}'$. If these quantum numbers are such that they allow a nonzero three-point function $f_{\calO\calO\calO}\sim\<\calO\calO\calO\>$
	then the beta-function for the perturbing coupling $\lambda_\perp$ will have the schematic form $-\eps \lambda_\perp+f_{\calO\calO\calO} \lambda_\perp^2$ where $\eps\sim x-x'$ and a real BZ fixed point will exist on both sides of $x'$. Suppose on the other hand that $\calO$ is odd under some $\bZ_2$ subgroup of $G'$, so that $f_{\calO\calO\calO}$ vanishes. The couplings $\lambda_\perp$ and $-\lambda_\perp$ are now equivalent and the beta-function has the form $-\eps \lambda_\perp+O( \lambda_\perp^3)$. Now fixed point at $x<x'$ lives at negative $\lambda_\perp^2$ which may e.g.~lead to violations of unitarity.}

Fig.~\ref{fig:scenario1} illustrates the considered non-minimal scenario. In it we see the line of BZ$'$ fixed points, of uncertain origin, which interchanges stability with BZ line at $x=x'$. For $x_c<x<x'$ the flow from free theory leads to BZ$'$ (and thus we have symmetry enhancement in IR to $G'$). At $x_c$ the line of BZ$'$ fixed points disappears by annihilation with a line of CFTs analogous to QCD${}^*$ (but with symmetry $G'$). The BZ line of CFTs with symmetry $G$ continues to exist at $x<x'$, but is unstable (contains a relevant $G$-singlet scalar). Being unstable, it does not affect the IR properties of RG flows originating in free theory at $x<x'$ and in principle it may continue to exist even for $x<x_c$.

Admittedly this scenario is a bit contrived, which is why we prefer the minimal scenario described in section \ref{sec:BZ}. However, it can eventually be probed by Monte Carlo simulations. Notice that in the non-minimal scenario approximate symmetry enhancement to $G'$ should appear even in the walking regime just below $x_c$. We will see in section \ref{sec:further} an example of an approximate symmetry enhancement without a fixed point, which may have a similar origin. It's easy to imagine even more complicated scenarios but we will stop here.

\section{Deconfined criticality: a further example of walking?}
\label{sec:further}

Here we will discuss an interesting 3d RG flow which, while not fully understood, seems to exhibit phenomena plausibly explainable by walking. This discussion is largely based on sections V.E.2 and V.E.4 of \cite{review} where further details and references can be found.

This RG flow has been originally brought up in condensed matter literature in relation with the N\'eel-Valence Bond Solid (VBS) transition and the phenomenon of  ``deconfined criticality" \cite{deconfined}.
Without going into condensed matter details, in field-theoretical language one studies the 3d Abelian Higgs model (also called bosonic QED${}_3$), which is the theory of a 3d $U(1)$ gauge field coupled to $N$ complex scalars $\phi_i$ with an $SU(N)$ invariant potential $m^2|\phi|^2+\lambda(|\phi|^2)^2$. For this discussion we focus on $N=2$ which is the most interesting and best-studied case, although $N>2$ is also interesting.

The global symmetry of this theory is $SO(3)\times U(1)_T$. The $SO(3)=SU(2)/\bZ_2$ part of the global symmetry descends from the $SU(2)$ symmetry of the scalar potential (one has to divide by $\bZ_2$ since the center part of $SU(2)$ corresponds to a gauge transformation). The $U(1)_T$ part of the global symmetry is topological, called $U(1)_T$ to distinguish it from the gauge group. The operators charged under it are $U(1)$ gauge field flux defects, called monopoles. 

In the above theory, by varying the UV mass $m^2$, one induces a phase transition between a Coulomb and a Higgs phase of the gauge field. This transition is supposed to describe the critical properties of the N\'eel-VBS transition in certain $(2+1)$d quantum antiferromagnets. One interesting question is whether this transition is continuous or first-order.

From Monte Carlo studies performed by various groups, the following picture transpires \cite{Nahum:2015vka,Nahum:2015jya}. First, no signs of a conventional first-order transitions are seen: it is either continuous, or perhaps a very weakly first-order (the correlation length being at least several hundred lattice spacings). Second, quite unexpectedly, near the phase transition the system is seen to possess a global symmetry enhancement from $SO(3)\times U(1)$ to $SO(5)$. For example, the N\'eel order parameter operator transforming as a fundamental of $SO(3)$ turns out to have the same scaling dimension $\Delta_\Phi=0.625(15)$ as the lowest charge monopole, which suggests that they may be combined into a $\Phi$ in the fundamental of $SO(5)$. For other impressive tests of $SO(5)$ enhancement see \cite{Nahum:2015vka,Nahum:2015jya}.

Assuming we have both a continuous phase transition and $SO(5)$ symmetry enhancement, one can ask about dimensions of lowest scalar operators $S$ and $T$ transforming as a singlet and a symmetric traceless tensor of $SO(5)$. The $T$ is known to be relevant of dimension $\sim 1.5$ \cite{Nahum:2015vka}, its various component being interpreted as the charge two monopole and the mass term $m^2|\phi|^2$ which drives the transition. On the other hand $S$ has to be irrelevant, because otherwise the transition would not be reached. 

Now, it turns out that this expectation comes into a clash with rigorous bounds computed from conformal bootstrap under the assumption of unitarity and $SO(5)$ invariance \cite{Nakayama:2016jhq,DSD2016}. Namely, these bounds imply that the irrelevance of $S$ requires $\Delta_\Phi>0.76$ \cite{Nakayama2016}, in conflict with the above measurement $\Delta_\Phi=0.625(15)$

The clash would be resolved if we assume that the phase transition is weakly first-order due to a walking behavior of the RG flow. The CFT controlling this flow being non-unitary and moreover complex, rigorous bootstrap bounds do not apply. The observed scaling exponents are then attributed to the approximately scale-invariant part of the RG trajectory. If this interpretation is correct, we are led to conclude that walking RG flow manages to relax the bound on $\Delta_\Phi$ quite significantly, from 0.76 for unitary theories down to the observed 0.625(15).

It's not our goal here to discuss various pro and contra in favor of this scenario. The hypothesis of a weakly first-order phase transition via walking mechanism was put forward in \cite{Nahum:2015jya} even before the above bootstrap evidence emerged, as one of the ways to explain some unusual finite-size scaling effects observed in their Monte Carlo simulations, and it was also discussed further in \cite{Wang:2017txt}. In \cite{part2} we will discuss peculiar form of deviations from scale invariance (drifting scaling dimensions) present in walking RG flows, possibly related to the unusual Monte Carlo effects seen in \cite{Nahum:2015jya}.

Finally, we mention two related condensed matter transitions which seem to exhibit similar physics.
First, there is the $N_f=2$ \emph{fermionic} QED${}_3$, for which the situation is as murky as for the N\'eel-VBS:
a continuous/weakly first-order dilemma, symmetry enhancement from $SU(2)\times U(1)$ to $SO(4)$, and a clash with bootstrap bounds, see section V.E.4 of \cite{review}. Second, there exists an easy-plane version of the N\'eel-VBS transition studied via numerical simulations in \cite{Zhao},\cite{Serna:2018tct}. In this case the situation is clearer: there is a weakly first-order phase transition due to walking behavior, as well as symmetry enhancement from $SO(3)\times \bZ_2$ to $SO(4)$. It would be interesting to understand better these examples.

Notice that in this paper we discuss walking flows with rotational invariance. This should be relevant for statistical physics examples of deconfined criticality. One may ask how our picture would be modified for quantum deconfined criticality, where the phase transition is driven by quantum fluctuations in anti-ferromagnets. The difference is that in this case we also expect scale-dependent deviations from rotational invariance, parametrized by `running of the speed of light'.\footnote{We are grateful to Silviu Pufu for raising this interesting point.} Running of the speed of light does not spoil our picture, but some of our RG computations need to be modified. When we perturb complex CFTs, as in section \ref{ssec:complexCFTS_walking}, we will have to add a marginal coupling to the $T_{00}$ component of the stress tensor into the RG equations. Interplay between this additional coupling and the walking coupling $\lambda$ will lead to new effects, but we believe our basic picture should be preserved. It would be interesting to work this out in detail.

\section{Complex CFTs}
\label{sec:Complex}

We have provided the reader with several examples of physical systems which show walking behavior. Both QCD with $x=N_f/N_c\lesssim x_c$ and the Potts model with $Q \gtrsim Q_c$ at the critical temperature have a large separation between the UV and the IR scale and a region of approximate scale invariance. As mentioned in section \ref{sec:walking}, RG behavior of walking systems is controlled by complex fixed points with small imaginary parameters. There we introduced complex fixed points to study the beta-function of the form \reef{eq:RGwalk}:
\be
\beta(\lambda)=-y-\lambda^2\,,
\label{eq:RGwalk6}
\ee
which for $y>0$ has fixed points $\calC,\bar{\calC}$ at complex values of the coupling constant $\lambda=\pm i\sqrt{y}$. At these fixed points the operator controlling the RG flow has anomalous dimension given in \reef{eq:Dpm}, namely 
\bea
&&\Delta=d-2i\sqrt{y},\qquad \lambda =i\sqrt{y}\,,\label{eq:C+}\\
&&\Delta=d+2i\sqrt{y},\qquad \lambda =-i\sqrt{y}\,.\label{eq:C-}
\eea
In this section we will proceed to study such complex fixed points in more detail.

We emphasize that each of these operators belongs only to the indicated fixed point but not to the other one. To have this feature it was crucial to consider an RG flow in the space of complex couplings.
To appreciate this last point better it is instructive to consider a different system of beta-functions for two real coupling constants:
\bea
&&\beta_u=2\sqrt{y}v\,, \nn\\
&&\beta_v=-2\sqrt{y}u \;.
\label{eq:RGuv}
\eea
This system of beta-functions may seem exotic, and indeed field theories that produce this kind of behavior are rather involved. In the classification which we will promulgate below they will count as real theories, albeit non-unitary. The motivation to call such theories real is that the coupling constants stay real, and moreover if one works in the basis of operators that correspond to couplings $u$ and $v$, correlation function are also manifestly real. We will present some examples below, but for now let us study these equations as an abstract toy model.

In case at hand there is a single fixed point for $u=v=0$ and there are two close-to-marginal operators at this fixed point with dimensions $d\pm2i\sqrt{y}$. In spite of the appearance of complex anomalous dimensions, which clearly indicates non-unitarity, the crucial difference of the fixed point of \eqref{eq:RGuv} from those of \eqref{eq:RGwalk6} is that both above complex-conjugate operators belong to it, while as mentioned 
each fixed point of \eqref{eq:RGwalk6} has a single complex operator. 

We will turn this distinction into a more formal statement in the next section, where 
we give some more details, definitions and examples of complex and real theories.

\subsection{Real vs complex QFTs} \label{sec:realandcomplex}

\subsubsection{RG evolution}

Let us start with the discussion of general QFTs and specify to CFTs later.\footnote{The reader interested primarily in complex CFTs can jump directly to section \ref{sssc:CFTs}.}
We would like to formalize the distinction between real and complex QFTs. It's best to proceed from examples. Consider e.g.~a perturbative Lagrangian theory of multiple real scalar fields. We can complexify coupling constants, considering them living in $\bC^M$ where $M$ is the total number of couplings. In this setup we can consider the subspace $\cal{R}=\bR^M$ of all couplings real. We would like to call theories corresponding to this subspace real. Notice that this subspace is preserved by RG evolution, so this looks like a natural definition.

For theories of real scalars in integer spacetime dimension $d$, the class of real theories coincides with that of unitary theories. However in general a real QFT does not have to be unitary. To see a 
simple example, let us couple scalars with vector fields, with all couplings real. The theory is still real, but as is well known it will be unitary only in a restricted class of theories respecting gauge invariance.

A general comment about complexifying RG evolution is in order. We assume that a coupling basis exists, such that beta-functions $\beta_a=dg_a/dt$ are locally analytic functions of complexified couplings with real coefficients: $\beta_a(\{g_b^*\})=\beta_a^*(\{g_b\})$. This guarantees that the subspace of real couplings $\calR$ is RG-preserved. Notice that it would be incorrect to think of the map $g_a\to g_a^*$ as \emph{simply} an example of a $\bZ_2$ symmetry under which the imaginary parts of all couplings are odd; it's much more than that. Of course we can split each coupling into real and imaginary part $g_a=\sigma_{a}+i \tau_{a}$ and view RG evolution as happening in $\bR^{2M}$. If we impose that $\tau_a$ are odd under a $\bZ_2$,  this would also explain why $\calR$ is preserved, but the above assumption of analyticity with real coefficients is much more constraining. Compare for example the beta-function $\beta(g)=g^2$ which in terms of real and imaginary parts reads
\beq
\beta_\sigma = \sigma^2-\tau^2\,,\quad \beta_\tau=2\sigma \tau\,.
\eeq
On the contrary imposing only $\bZ_2$ would allow quadratic beta-functions of the same functional form but with arbitrary relative coefficients.

\subsubsection{Correlation functions}

Definition of real vs complex theories in terms of RG evolution is intuitively clear, 
but we would like to have a definition that may be applicable to theories which contain fields that are intrinsically complex, as well as theories which do not necessarily admit a Lagrangian description. Such a definition can be given in terms of correlation functions: we will call a theory real if it contains a set of operators $\calO_i$ whose correlation functions are real at all distances.

A slightly more nuanced but practically almost equivalent definition is as follows. For simplicity let us focus on theories which have parity invariance and let's talk only about bosonic operators. The theory is called real if there is an involutive map $*$ which acts on local operator labels and puts in correspondence to each local\footnote{Here as in the rest of the paper we focus for simplicity on local operators, however the conjugation relation in a real theory should exist also for non-local operators. We thank Silviu Pufu for inquiring.} operator $A_i$ an operator $A^*_i$ such that the correlation functions of $A$'s and $A^*$'s, while in general complex, are complex conjugate of each other:
\beq
\langle A_1(x_1) A_2(x_2)\ldots A_n(x_n)\rangle^*=	\langle A^*_1(x_1) A^*_2(x_2)\ldots A^*_n(x_n)
\rangle\,.
\label{eq:AB}
\eeq
Hopefully it will not be confusing that we use the same symbol $*$ as a complex conjugation acting on numbers, as well as a map acting on names of operators. Real operators are those whose correlation functions are real, and so according to the above definition we have $A^*=A$ for such operators.\footnote{\label{note:ReIm}For any $A$ the operators $A+A^*$ and $i(A-A^*)$ will be real.} 

On the contrary, if the map $*$ with the above properties does not exist, then the theory is classified as complex.

Notice that the above definition makes sense separately of any quantization interpretation. So the operation $*$ does not have to be thought of as a conjugation of operators acting in some Hilbert space. If we know all correlation functions of the theory, we can inspect them and decide if the map $*:A\mapsto A^*$ exists. 

However if we do have a parity-invariant unitary theory realized in a Hilbert space, then it's easy to see that it would be classified as real according to the above definition with $\calO^*=(-1)^{p_\calO} \calO^\dagger$ where $p_\calO$ is the parity of $\calO$. 
Let us split $x=(\tau,\mathbf{x})$ and use quantization by planes in the $\tau$ direction, so that 
$\calO(\tau,\mathbf{x})^\dagger = \calO^\dagger(-\tau,\mathbf{x})$ and so
\begin{equation}
\< \calO_1 (x_1) \ldots \calO_n(x_n)\>^*
= \< \calO_1^\dagger (-\tau_1,\mathbf{x}_1) \ldots \calO_n^\dagger (-\tau_n,\mathbf{x}_n)\>\,.
 \label{eq:conj3}
\end{equation}
Using parity transformation we can now flip all $\tau$'s and go back to the equation of the form 
\reef{eq:AB} where the operators in the r.h.s.~and l.h.s.~are at the same positions.

So, all unitary theories are real but of course unitary theories in Euclidean space satisfy a crucial additional assumption, the reflection positivity, which is the positivity constraint on $(2n)$-point functions:
\begin{equation}
\langle \calO_1^\dagger(-\tau_1,{\bold x_1}) \calO_2^\dagger(-\tau_2,{\bold{x_2}})\ldots \calO_2(\tau_2,\bold{x_2}) \calO_1(\tau_1,\bold{x_1})\rangle \ge 0\,. \label{eq:reflpos}
\end{equation}
We refer the reader to e.g.~\cite{Haag:1992hx} for precise definition of reflection positivity. Reflection positive theories can be analytically continued to Minkowski space in a consistent way.

One the space of complex QFT's it is natural to define the complex conjugation map such that for any operator $A_i$ present in the original theory the complex conjugate theory contains operator $A_i^*$ and correlation functions of the operators in two theories are related by Eq.~\eqref{eq:AB}. Then real theories are the fixed points of the conjugation map.

\subsubsection{Examples}
\label{sec:exQFT}

Let us now give some examples of real yet non-unitary Euclidean theories. One example was already mentioned: theories of multiple scalars and vectors without gauge invariance, coupled with real couplings.

A more subtle example is a real scalar $\phi$ with potential $V(\phi)= i h \phi +i \lambda \phi^3$. This potential satisfies $V(-\phi)=V(\phi)^*$. This theory will be real according to the correlation function definition, with $\calO=i\phi$ being a real operator.\footnote{Such theories are sometimes called PT-invariant. Literature on PT-invariant theories is sometimes hard to read because valid results on Euclidean PT-invariant theories are often interspersed with highly suspect claims that such theories may somehow be relevant also for real-time, Lorentzian physics, in spite of being non-unitary.} We will encounter the IR fixed point of this theory in section \ref{sec:exCFT} as a real but non-unitary CFT -- the Lee-Yang minimal model $\calM_{2,5}$.
 
As the next example, consider $O(N)$ models analytically continued to non-integer $N$. Correlation functions stay real, at least in perturbation theory, but these theories are non-unitary \cite{Maldacena:2011jn}.

Finally, consider theories of real scalars with real couplings, analytically continued to non-integer Euclidean dimensions $d$, {\it \`a la} Wilson-Fisher. Such theories have been shown to be non-unitary in \cite{Hogervorst:2014rta,Hogervorst:2015akt}. It would be nice to clarify to which extent they are nonperturbatively well-defined.

\subsection{Real vs complex CFTs} 
\label{sssc:CFTs}
We now proceed to discuss real and complex CFTs. Since the structure of CFTs is more constrained we will be able to make our definitions more concrete. Real (complex) CFTs can arise as fixed points of real (complex) RG flows. We will discuss the consequences of reality on the spectrum of a CFT, and provide the reader with some examples.

A CFT is defined by its conformal data: the set of all operator dimensions and all OPE coefficients. 
Following the discussion of the previous section, if a CFT is real, and an operator $\calO$ with complex scaling dimension $\Delta$ is part of the theory, then also $\calO^*$, with scaling dimension $\Delta^*$, must be part of the theory. Some operators will have real scaling dimensions, while operators with complex scaling dimensions can exist only in complex conjugate pairs in order for $\calO^*$ to exist.\footnote{Notice that when dealing with a real \emph{QFT}, we could always pass from any operator to its real and imaginary part (footnote \ref{note:ReIm}) which are real operators. For real \emph{CFTs} this is not a natural thing to do, because if $\calO$ is an operator of complex scaling dimension, its real and imaginary part will not have a well-defined scaling dimension.}

OPE coefficients must satisfy relations that follow from \eqref{eq:AB} applied to three-point functions, in particular OPE coefficients of three real operators must be real.\footnote{Note that in unitary CFTs OPE coefficients are known to satisfy reality constraints, see \cite{review}. Here we are describing a context when it is natural to impose reality of OPE coefficients even if the CFT is not unitary.} 

If instead we consider a complex CFT, operators with complex scaling dimension can appear without their complex conjugated partner being present in the theory. Similarly OPE coefficients can be complex numbers not subject to any obvious constraints. Central charge of complex CFTs can be a complex number, as we will see in \cite{part2}. Despite the fact that conformal data is complex, complex CFTs still fulfill other usual properties: conformal symmetry, operator product expansion, and crossing. 

\subsubsection{Examples}
\label{sec:exCFT}

We will now review some examples. 
As we saw above, all unitary CFTs are real. Let us consider examples of real but non-unitary CFTs, in order to highlight the difference between reality and unitarity. 

Consider first 2d examples. The simplest example is the Lee-Yang minimal model $\calM_{2,5}$ which appears as an IR fixed point of a theory of 2d scalar with purely imaginary cubic coupling \cite{Fisher:1978pf, Cardy:1985yy}, see section \ref{sec:exQFT}.
This CFT has real spectrum, with a single Virasoro primary $\phi$ of dimension $h=\bar h=-1/5$ and real central charge $c=-22/5$. That $h$ and $c$ are negative is a clear sign of non-unitarity. In the usual normalization the OPE coefficient $C_{\phi\phi\phi}$
is purely imaginary. The real field is $\tilde \phi=i\phi$, with a real OPE coefficient $C_{\tilde\phi\tilde\phi\tilde\phi}$. This CFT is thus real non-unitary.

The previous example generalizes to all non-unitary minimal models $\calM_{p,q}$. Recall that $\calM_{m,m+1}$ are unitary while for $|p-q|>1$ the minimal models $\calM_{p,q}$ are non-unitary.
We consider integer $p,q$ so that there is a finite Kac table and a finite number of primaries. In spite of being non-unitary, all primary fields in these theories have real scaling dimensions $h_{r,s}$ and the central charge is real. The OPE coefficients in minimal models were investigated by Dotsenko and Fateev \cite{Dotsenko:1984nm,Dotsenko:1984ad}. It follows from their work that pairwise products of OPE coefficients $C_{ijk}C_{klm}$, for which they give explicit formulas, are real. This means that either all OPE coefficients are real or purely imaginary, in which case they are made real multiplying all fields by $i$. So these non-unitary minimal models are real CFTs.

Continuing the list of 2d examples, critical Potts model with $Q$ non-integer and $Q<Q_c$ will be real but non-unitary. That they are non-unitary can be seen very easily from their central charge and the spectrum, which are exactly known as we will discuss in \cite{part2}. Also the random cluster measure which is the microscopic origin of these critical point is known not to have reflection positivity \cite{Biskup1998}. On the other hand the random cluster description is manifestly real, and so it guarantees that the critical point if it exists should be a real CFT.

The same holds for the $O(n)$ model with non-integer $n$ in 2d, which can be given a nonperturbative microscopic definitions as theories of loops, and is known to have a critical point for $n\in[-2,2]$. Much is known about these CFTs, see e.g.~\cite{diFrancesco:1987qf}. These are also examples of real non-unitary CFTs.

These examples have purely real conformal data and can be hardly confused with complex theories. In this regard it is useful to bring up a CFT which is still real, but has pairs of complex conjugate operator dimensions. It turns out that Wilson-Fisher fixed point in $d=4-\eps$ dimensions is a theory of this sort \cite{Hogervorst:2015akt}.\footnote{See also \cite{Hogervorst:2014rta} for prior work and \cite{DiPietro:2017vsp} for a related fermionic example.}
The Wilson-Fisher fixed point in $d=4-\eps$ is a textbook example of a weakly coupled fixed point. It is described by a massless boson perturbed by a quartic interaction term with a real coupling. In $d=4$ the theory is unitary: all states have positive norm; however, when we move to $d=4-\eps$ the situation changes. There are some \textit{evanescent operators} which have zero norm in integer dimension, but can have negative norm in fractional dimension: it follows that the theory at the Wilson-Fisher fixed point is non-unitary. This is also reflected in the spectrum of the theory: because of the negative norm states it is possible for some operators to acquire a complex anomalous dimension. This happens for some of these evanescent operators, and complex scaling dimensions always appear in complex conjugate pairs, as expected in a real CFT. The existence of these complex operators was a bit hard to notice, since at first-order in $\eps$ they appear at very high dimension ($\Delta=23$) \cite{Hogervorst:2015akt}. 

Another curious physical example of a real theory with pairs of complex operator dimensions is a long-range disorder fixed point studied in \cite{Halperin}. 

Continuing with higher-dimensional examples, it is worth mentioning that the Lee-Yang CFT described above can be studied for any $2\le d< 6$, as a fixed point of the cubic scalar theory with imaginary coupling, and we expect it to be real for all $d$ in this range. Close to the upper critical dimension, in $d=6-\eps$, the theory can be studied perturbatively. It would be interesting to see if the spectrum of the theory is completely real in $d=6-\eps$ as it is in $d=2$, or if operator pair with complex conjugate dimensions occur.

Finally let us discuss examples of complex CFTs. These examples are less frequent in the literature than real non-unitary theories, and there seems to be no general consensus if they are physical and/or well defined. Our first example is $\calN=4$ SYM: since the beta-function vanishes for all values of $g^2$, it appears that if we give $g^2$ an imaginary part, we should obtain a complex CFT.\footnote{In the planar limit of $\calN=4$ SYM, scaling dimensions of operators appear to be analytic functions of the 't Hooft coupling $\lambda$, $\lambda^2=g^2N_c/(16\pi^2)$, in the disk $|\lambda|<1/4$ (see e.g.~\cite{Leurent:2013mr}, Eq.~(67)). We thank Carlo Meneghelli and Pedro Vieira for discussions.} Notice that this procedure is fundamentally different from considering the theory as a function of the complexified couplings $\tau,\bar\tau=\frac{\theta}{2\pi}\pm\frac{4\pi i}{g^2}$, with $\theta$ the theta-angle. Taking $g^2$ complex means that we are considering the situation when $\bar\tau\ne \tau^*$.\footnote{Note in this respect the limit $\bar\tau\to -i \infty$ with $\tau$ fixed considered in \cite{Frenkel:2006fy}, although there it was interpreted as taking $g\to0$ and theta-angle imaginary and large. We thank Nikita Nekrasov for discussions.}  More generally, we can consider a supersymmetric CFTs possessing a set of exactly marginal couplings $\lambda_i$, and consider it as a function of complexifed couplings $\lambda_i\in \bC^M$, which gives rise to complex CFTs. Partition functions for such complex deformations were discussed in \cite{Gerchkovitz:2014gta}. On the other hand Ref.~\cite{Bobev:2016nua} considered SCFTs perturbed by complex mass deformations, which in our language corresponds to complex QFTs.\footnote{We thanks Silviu Pufu for discussions.}
 
{In studies of RG flows of multi-scalar theories using the $\eps$-expansion and $1/N$-expansion, it's not unusual to encounter RG fixed points at complex couplings, which should be interpretable as complex CFTs living in non-integer dimension $d$. Many such examples have appeared recently in the work of Simone Giombi, Igor Klebanov and collaborators \cite{Fei:2014yja,Fei:2014xta,Giombi:2017dtl,Giombi:2018qgp}, with complex fixed points arising from real fixed points which annihilate and go into the complex plane when varying the number of fields $N$ or the dimension $d$. While these works view complex operator dimensions as a sign of instability, and refer to complex fixed points as ``unstable CFTs", more optimistically these theories could actually be nonperturbatively well-defined Euclidean CFTs.}
  
 A further example are the fishnet theories \cite{Gurdogan:2015csr}, obtained as deformations of $\calN=4$ SYM at large $N$ and in some special double limit of the coupling and of the twists. These deformations break supersymmetry, and the fishnet theories are non-supersymmetric fixed points with complex anomalous dimensions and no pair of complex conjugate operators \cite{Grabner:2017pgm} --- they appear to be complex theories according to our definition.

Our final example is the complex fixed point for the Potts model with $Q>Q_c$, to be studied in detail in \cite{part2}. It allows for many explicit computations which significantly clarify the concept of complex CFT.

\subsection{Complex CFTs and walking} 
\label{ssec:complexCFTS_walking}
We have seen in section \ref{sec:walking} that the walking beta-function has two zeros at complex coupling, and walking behavior of a theory can be understood as the flow passing in between these complex fixed points, when they are close to the real axis. Now we would like to reverse the logic and show how, by starting from a pair of CFTs in the complex plane of the coupling, we can describe the real walking theory. From the practical point of view, this section is perhaps the most important one in this paper: while the rest of our work was devoted mostly to clarifying misunderstandings (some of the them our own) and laying conceptual foundations, here we will propose a concrete calculational 
scheme.

Let us start with two complex theories, $\calC$ and \cC, which are related by complex conjugation, meaning that the spectrum of \cC$ $ is the complex conjugate of the spectrum of $\calC$. A similar condition holds for OPE coefficients. 
We will formalize the condition that these two theories are close to being real, by requiring that scaling dimension and OPE coefficients have an imaginary part of order $O(\eps)$, where $\eps\ll 1$. We will assume that this condition should hold at least for low-lying operators.\footnote{As we said before, a general real theory could have operators with complex scaling dimension, provided that they appear in complex conjugate pairs. We are assuming here that $\calC$ and $\barcalC$ are close to a more restricted real theory, where all operator dimensions are real, at least in the low-dimension part of the spectrum.} Furthermore, we assume that the spectrum of $\calC$ contains one almost marginal operator $\calO$ of dimension $\Delta_{\calO}=d - i \eps+O(\eps^2)$, which is a singlet under the global symmetry group. We emphasize once again that the operator of the complex conjugate dimension then belongs to $\barcalC$, and is not a part of $\calC$.

We will develop a form of conformal perturbation theory (CPT) where we perturb $\calC$ by $\calO$. Usually, CPT is used to describe RG trajectories which either flow out from a CFT or flow into it. The difference here is that we will use CPT to describe an RG trajectory which approaches CFT but does not necessarily touch it, as in Fig.~\ref{fig:fig2}.
Apart from this difference of interpretation, formally we proceed as usual in CPT, perturbing the action of $\calC$ by adding operator $\calO$ with some (in general complex) coupling $g$:
\beq
S_{\calC}+ g \int d^d x\, \calO(x)\,.
\eeq
For nonzero $g$, the scale invariance of the theory is in general broken and the coupling $g$ will run. The one loop beta-function is given by the standard result (see e.g.~\cite{Cardy:1996xt})
\beq
\beta_g=-i \eps g +\frac{1}{2}S_dC_{\calO \calO \calO} g^2+\ldots\,, \label{eq:betaCPT}
\eeq
with $S_d$ the volume of the $d$-dimensional unit sphere and $C_{\calO \calO \calO}$ the OPE coefficient extracted from the three-point (3pt) function $\langle \calO \calO \calO \rangle$. In general $C_{\calO \calO \calO}$ is complex, but at the order we will be working here, we can neglect its $O(\eps)$ imaginary part and treat it as real.

We see that the above beta-function has two fixed points: the trivial $g=0$, and the nontrivial at $g=g_{FP} =i \frac{2 \eps}{S_d C_{\calO \calO \calO} }$, which we denote $\calC_{FP}$. Since there is only one almost marginal singlet operator, and $\calC$ and \cC$ $ are close to each other, it appears reasonable to identify $\calC_{FP}$ with $\barcalC$. One simple check of this identification is to compute the scaling dimension of the operator $\calO$ at this fixed point:
\beq
[\calO]_{g=g_{FP}}=d+\beta_g'(g_{FP})=d+i \eps+O(\eps^2)\,.
\eeq
We see that the imaginary part flipped sign as expected.

What about the other operators? Along the flow, a generic operator $\phi$ acquires an anomalous dimension (see e.g.~\cite{Cardy:1996xt})
\beq
\gamma_\phi(g)=S_d C_{\phi \phi\calO} g + O(g^2)\,.
\eeq
Its scaling dimension at the fixed point is $\Delta^{\calC}_\phi+\gamma_\phi(g_{FP})$. To identify it with the dimension of the same operator at $\overline\calC$, the imaginary part must flip sign (we call it the Im-flip condition).
For operator $\calO$ this happened automatically because of the way the coefficient $C_{\calO\calO\calO}$ controlled the calculation, but for a generic operator this requires a non-trivial relation between OPE coefficients and scaling dimensions. At one loop we should have (all quantities refer to CFT $\calC$):
\beq
\frac{\rm Im \Delta_\phi}{C_{\phi \phi\calO}}=\frac{\rm Im \Delta_\calO}{C_{\calO \calO \calO}}\,, \label{eq:OPEcondition}
\eeq
up to corrections higher order in $\eps$. 

Here's an intuitive argument in favor of \reef{eq:OPEcondition}. Assume for a second that it does not hold for some operator $\phi$. Then the dimension of $\phi$ at the fixed point will be different from $\Delta_\phi^*$, and hence $\calC_{FP}$ cannot coincide with \cC. So assuming such a scenario, we have four nearby complex CFTs: $\calC$, $\barcalC$, $\calC_{FP}$, $\overline{\calC_{FP}}$, as opposed to only $\calC$ and $\barcalC$. This proliferation of CFTs seems rather unlikely. It is more economical to assume that in fact $\calC_{FP}=\barcalC$, which requires \reef{eq:OPEcondition}. Hopefully in the future \reef{eq:OPEcondition} will be derived from general CFT principles (like OPE and crossing) applied to the pair of complex CFTs, although at the moment we don't have such a proof. In \cite{part2}, we will give an explicit example where this relation is satisfied by several operators. 

Now we would like to recover the real walking theory. Intuitively, it corresponds to the RG trajectory which passes in the middle between the two complex CFTs. We should land on this trajectory by adding half of the coupling that takes us from $\calC$ to \cC. Adding the operator $\calO$ with a coupling $g=\frac{g_{FP}}{2}-\lambda$, the above beta-function re-expressed in terms of $\lambda$ takes the form:
\beq
\beta_\lambda=-\frac{\eps^2}{2S_d C_{\calO \calO \calO}}-\frac{S_d C_{\calO \calO \calO}}{2}\lambda^2 +\ldots \label{eq:betaf}
\eeq
We see that to the considered order all imaginary terms cancel: the coupling $\lambda$, if it starts real, will remain real during the RG evolution. Rescaling $\lambda$, we bring the beta-function to the walking beta-function \eqref{eq:RGwalk} with $y=\eps^2/4$. Since $y$ does not depend on $C_{\calO\calO\calO}$, the generated hierarchy \reef{eq:xiWalk} in this one-loop order is independent of the OPE coefficient.

To further check that the theory described by the flow \eqref{eq:betaf} is indeed real, we should compute correlation functions of local operators and show that they are real. For 2pt function this will be done in \cite{part2}, where we will also compute deviations from scale invariance interpreted as ``drifting scaling dimensions".

Finally, the following comment on the validity of CPT is in order. To land on the real trajectory, we need to perturb with the coupling $g_{FP}/2$, which is proportional to the imaginary part of the dimension of $\calO$. Consequently, {the latter} needs to be small in order for the perturbation theory to be under control. Higher order calculations performed in \cite{part2} will demonstrate that it is actually the square of the imaginary part that serves as an expansion parameter for real physical quantities, which somewhat improves the convergence properties of CPT.

To summarize, in this section we sketched a technique whereby, once the conformal data of the complex CFTs is known, CPT can be used to make predictions for correlation functions in the walking regime. We limited ourselves to the leading-order CPT for simplicity, but one can go to higher orders provided that the conformal data of the complex CFT is known to $O(\eps^2)$ or even exactly. To put the perturbative computation under control (in the walking regime), it was convenient to assume that there is a family of CFTs depending on the continuous parameter $\epsilon$. If instead the complex CFTs are isolated, meaning that there is no such continuous parameter, one can still do an expansion provided that 
$(\Delta_{\calO}-d)^2$ is numerically small.\footnote{For subtleties related to such an expansion see the recent discussion in section 3.4 in \cite{Komargodski:2016auf}.}

\section{Conclusions}
\label{sec:conclusions}

To conclude, let us briefly summarize the main results of this paper. We introduced a new type of conformal field theories that we call complex since they correspond to fixed points of RG flows that exist at complex values of coupling constants. We argued that these theories can be well defined, and that one can work with them in the same way as one does with usual real CFTs. Importantly, these complex fixed points actually control RG flows of some real and unitary \emph{gapped} physical theories. This happens if the parameters of a complex CFT have small imaginary parts and the real RG flow, which we actually are interested in, is forced to pass close to it. As usual, in the proximity of a fixed point the RG flow becomes slow which leads to various interesting phenomenological properties, like a large hierarchy of scales and a large correlation length. The corresponding RG behavior is referred to as walking.

Examples of applications of complex CFTs include certain gauge theories near the end of the conformal window, as well as various condensed matter systems that exhibit weakly first-order phase transitions. Previous approaches to describing these systems, however, centered their discussion around real fixed points that exist only if certain parameters of the theory are altered. We claim that a better-controlled approximation and more physical understanding arises with our method. In this paper we focused mostly on drawing the general physical picture, and showed how to back it up with elementary computations. Further evidence will be provided in a companion paper \cite{part2} in which we study a very clean and amenable to detailed calculations example of walking RG --- the two-dimensional Potts model with number of states $Q$ larger than four. There we construct explicitly the corresponding complex CFT and check that its perturbation indeed describes the model of interest.

We reckon that thinking in terms of complex CFTs will improve our understanding of various models studied in high energy particle physics, as well as of condensed matter systems.

\section*{Acknowledgements}

We are grateful to Ignatios Antoniadis, Alexei Borodin, John Cardy, John Chalker, Matthijs Hogervorst, Sergei Gukov, Jesper Jacobsen, Elias Kiritsis, Maxim Kontsevich, Carlo Meneghelli, Adam Nahum, Nikita Nekrasov, Eliezer Rabinovici, Hubert Saleur, Senthil Todadri, Balt van Rees, and Pedro Vieira for useful discussions and comments, and to Bernard Nienhuis for communications concerning \cite{NienhuisPRL}. We thank Silviu Pufu and Marco Serone for comments on the draft. SR is grateful to Jesper Jacobsen whose lectures at CERN in 2015 got him interested in the Potts model, and to Damon Binder for enlightening discussions about category theory. VG is grateful to CERN and ENS for hospitality. We are all grateful to Caltech for hospitality during the completion of this work, and to the organizers and participants of ``Bootstrap 2018" for their interest and comments.

SR is supported by the Simons Foundation grant 488655 (Simons Collaboration on the Nonperturbative Bootstrap), and by Mitsubishi Heavy Industries as an ENS-MHI Chair holder. BZ is supported by the National Centre of Competence in Research SwissMAP funded by the Swiss National Science Foundation. 

\appendix

\section{Tuning and weakly first-order phase transitions}
\label{sec:1st-order}

First-order phase transitions have several characteristics, but the one which will be most useful to us is that the correlation length $\xi$ for fluctuations of the order parameter remains finite at such a transition.
Some first-order transitions are classified as weak. Once again there are various characterizations of what this means. A useful for us definition is that the correlation length at a weak first-order (WFO) phase transition becomes very large with respect to the microscopic length scale (e.g.~the lattice spacing): $\xi\gg a$.

Continuous phase transitions, which have $\xi=\infty$, are understood by RG theory as nontrivial fixed points of RG flow,\footnote{We call by trivial the fixed points describing the ordered and disordered phases with finite correlation lengths.} usually described by CFTs. WFO transitions are in a sense ``almost continuous", and one expects the RG theory to say something about them.
An RG flow trajectory which corresponds to a WFO transition is long (so that a hierarchy $\xi\gg a$ results), but it does not lead to a CFT fixed point, because otherwise the transition would be continuous.

How such an RG trajectory can arise? Walking is one possibility. 
However, weakness of some first-order phase transitions is explained via the tuning scenario, as we will now review.

\begin{figure}[h]
	\begin{center}
		\includegraphics[scale=.3]{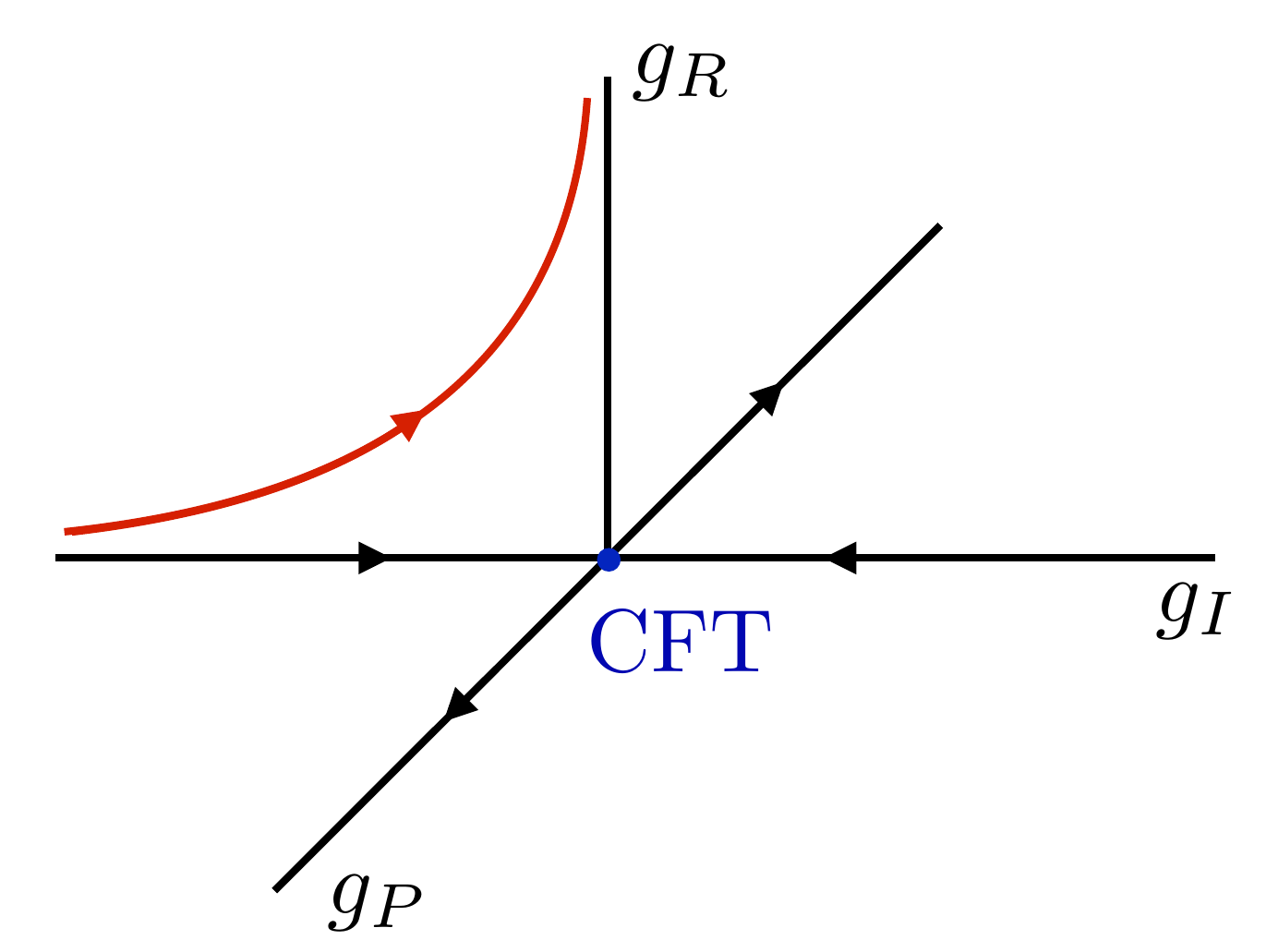}
		\caption{The tuning mechanism for a weakly first-order phase transition.}
		\label{fig-flyby}
	\end{center}
\end{figure}

In this scenario a CFT fixed point exists, but an RG flow near-misses it because of an extra relevant coupling turned on, see Fig.~\ref{fig-flyby}. Here we see a flow in the space of three couplings, $g_P$, $g_R$ (both relevant) and irrelevant $g_I$, which play different physical roles.
We assume that $g_P$ is the control parameter which is tuned in the UV to reach the transition. As such it's a relevant perturbation of the CFT. The $g_R$ represents another relevant perturbation of the CFT, while $g_I$ stands collectively for all irrelevant perturbations. 

If $g_R=0$ in the UV, then by tuning $g_P$ we can get onto a trajectory which ends in a CFT --- this would be a continuous transition. Suppose instead that our system has a nonzero value of $g_R$ in the UV. Then we end on up on the red trajectory which misses the CFT. That trajectory may end up at another CFT (in which case the transition is still continuous but in a different universality class). But it might as well happen that the red trajectory ends in a gapped theory --- then the transition is first-order.\footnote{We assume that the trajectories for $g_P<0$ and $g_P>0$ flow to two different phases, otherwise there is no phase transition to talk about.} 

Now imagine that we tune $g_R$ to be very small in the UV. Then the RG trajectory will spend a lot of RG time near the CFT, until eventually ending in a gapped phase. In this case the first-order transition will be weak. The RG flow duration depends on how strongly relevant the $g_R$ perturbation is near the CFT. If the scaling dimension of that perturbation is $\Delta_R<d$, then by the standard RG reasoning we expect the hierarchy
\beq
\label{eq:xia}
\xi/a \sim g_R^{-1/(d-\Delta_R)},
\eeq
where we express $g_R$ in dimensionless units at the lattice scale. This way of generating a WFO phase transition is precisely the tuning mechanism from section \ref{sec:Tuning}, see Eq.~\reef{eq:hier}.

As in section \ref{sec:Tuning}, there are two ways to get a very large number in the r.h.s. of \reef{eq:xia}:
\begin{enumerate}
	\item 
If $d-\Delta_R=O(1)$, we have to take $g_R$ very small. 
\item 
If, on the other hand, our CFT has a property that $d-\Delta_R\ll 1$, then it's enough to take $g_R$ somewhat smaller than one. E.g. if $d-\Delta_R=0.25$ and $g_R=0.25$ we get $\xi/a\sim256$ which starts being a large number. This latter possibility is what we called `mild tuning' in section \ref{sec:Tuning}.
\end{enumerate} 

To illustrate possibility 1, consider the Ising model, for definiteness in 3d, in a small nonzero magnetic field. We 
pick $g_P$ magnetic field, $g_R$ deviation from the critical temperature $T_c$. If $g_R=0$, then for $g_P=0$ we have a second order transition, governed by the critical 3d Ising CFT. The leading operator which couples to $g_R$ is the energy operator $\vareps$, of dimension $\Delta_R\approx1.41$. For $g_R<0$ the transition becomes first-order. Since $d-\Delta_R=O(1)$, the transition is weak only if $g_R$ is very small. 

To illustrate possibility 2, consider the CFT $\calC_0$ consisting of $N$ decoupled critical 3d Ising CFTs. The global symmetry of this fixed point is the \emph{cubic group}, generated by independent $\bZ_2$ transformations of each copy and by permutations of the copies. We are interested in RG flows which preserve this symmetry at the microscopic level. There are only two relevant singlet operators:
\beq
\calO_P=\vareps_1+\ldots+\vareps_N,\qquad
\calO_R=\sum _{i<j} \vareps_i \vareps_j\,,
\eeq
where $\vareps_i$ are the energy perturbations of each Ising CFT.\footnote{We constructed this example inspired by \cite{Zohar}.} The dimension of $\vareps$ being $\approx1.41$, we have
$\Delta_R\approx 2.82$. We get a transition varying $g_P$. Consider the nature of the transition as a function of $g_R$. If $g_R=0$ the transition is continuous. Consider what happens if $g_R\ne 0$, depending on the sign. If $g_R>0$, the RG trajectory misses the CFT $\calC_0$. It is known that in that case it leads to another CFT, which is the critical $O(2)$ model for $N=2$, and the so-called ``cubic fixed point" for $N\ge 3$. If $g_R<0$, the trajectory also misses $\calC_0$ and leads to a gapped phase --- this will be a first-order transition. Since $d-\Delta_R\ll 1$, a moderate tuning in $g_R$ will suffice for systems with this symmetry to exhibit a \emph{weakly} first-order phase transition.

The just given example is important for understanding why some antiferromagnets with multicomponent order parameters exhibit first-order phase transitions \cite{Binder,Pelissetto:2000ek}. This first-order phase transition is referred to in the literature as ``fluctuation driven", for the following reason. In the Landau-Ginzburg description, we would describe the above RG flow in terms of a multicomponent scalar theory with the mass term 
\beq
m^2(\varphi^2_1+\ldots+\varphi_N^2)=m^2 \vec\varphi^{\,2}
\eeq
and two quartic interactions allowed by the cubic symmetry
\beq
u(\vec\varphi{\,^2})^2+v(\varphi^4_1+\ldots+\varphi_N^4)\,.
\eeq
The decoupled fixed point has $u=0$, $v=v_0>0$ and some critical value of $m^2$. Perturbing by $g_R$ corresponds to turning on a small $u$. If $|u|\ll v_0$, of whatever sign, the quartic potential is stable, and the Landau theory predicts a second-order phase transition. However, RG analysis, which is under perturbative control in $d=4-\eps$ dimensions, shows that $u<0$, however small, grows more negative so that eventually the flow leads to un unstable potential, with the conclusion that the transition is actually first-order. This is the origin of the ``fluctuation driven" terminology, included here for historical reasons. If one thinks nonperturbatively, this terminology does not play much of a role.

\section{Walking vs BKT transition}
\label{sec:BKT}

Let us briefly review the BKT transition and explain why, while superficially it shares some similarity with the basic scenario of walking as presented in section \ref{sec:hierarchy}, there are also some important differences of principle.

BKT-like transition arises when three conditions are satisfied:
\begin{itemize}
	\item there is a one-parameter family of CFTs $\calT(K)$ related by an exactly marginal deformation (coupling) $K$, singlet of the global symmetry group. Let's call the corresponding singlet scalar operator $\calO_0$, of dimension $\Delta_{0}=d$ for any $K$;
	\item along this family, there is another singlet scalar operator, call it $\calO_1$, whose scaling dimension $\Delta_1(K)$ varies monotonically with $K$ and crosses from relevant to irrelevant at $K=K_c$. 
		\item the leading nontrivial operator which occurs in the OPE $\calO_1\times \calO_1$ is $\calO_0$, so that the OPE coefficient $C_{110}\sim \<\calO_1 \calO_1 \calO_0\>$ is nonzero. 
\end{itemize}

In the original BKT transition, as reviewed e.g.~in \cite{Cardy:1996xt}, we have $d=2$ and the family $\calT(K)$ is the massless scalar boson $\theta$ compactified on $[0,2\pi]$ and with the action $\half K \int d^2x\, (\nabla \theta)^2$. So $\calO_0$ is just the operator multiplying $K$ in the action. It is the singlet of the global $U(1)$ symmetry of the CFT.\footnote{More precisely, the global symmetry is $U(1)_L\times U(1)_R$.} On the other hand $\calO_1$ is the vortex operator which inserts a defect, its scaling dimension being $\Delta_1(K)=\pi K$, and $K_c=2/\pi$.\footnote{More precisely, $\calO_1$ is the sum of the charge-1 vortex and antivortex operators.} The OPE coefficient $C_{110}$ is indeed nonzero in this case.

\begin{figure}[htbp] 
	\centering
	\includegraphics[scale=.4]{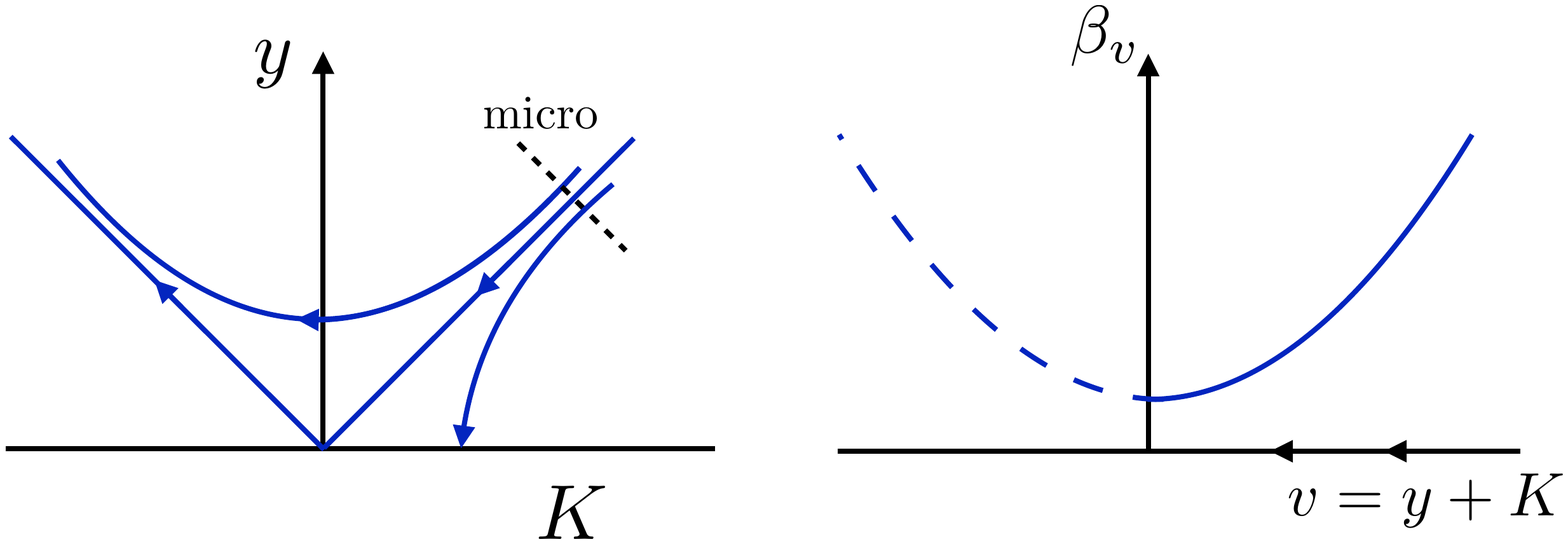} 
		\includegraphics[scale=.4]{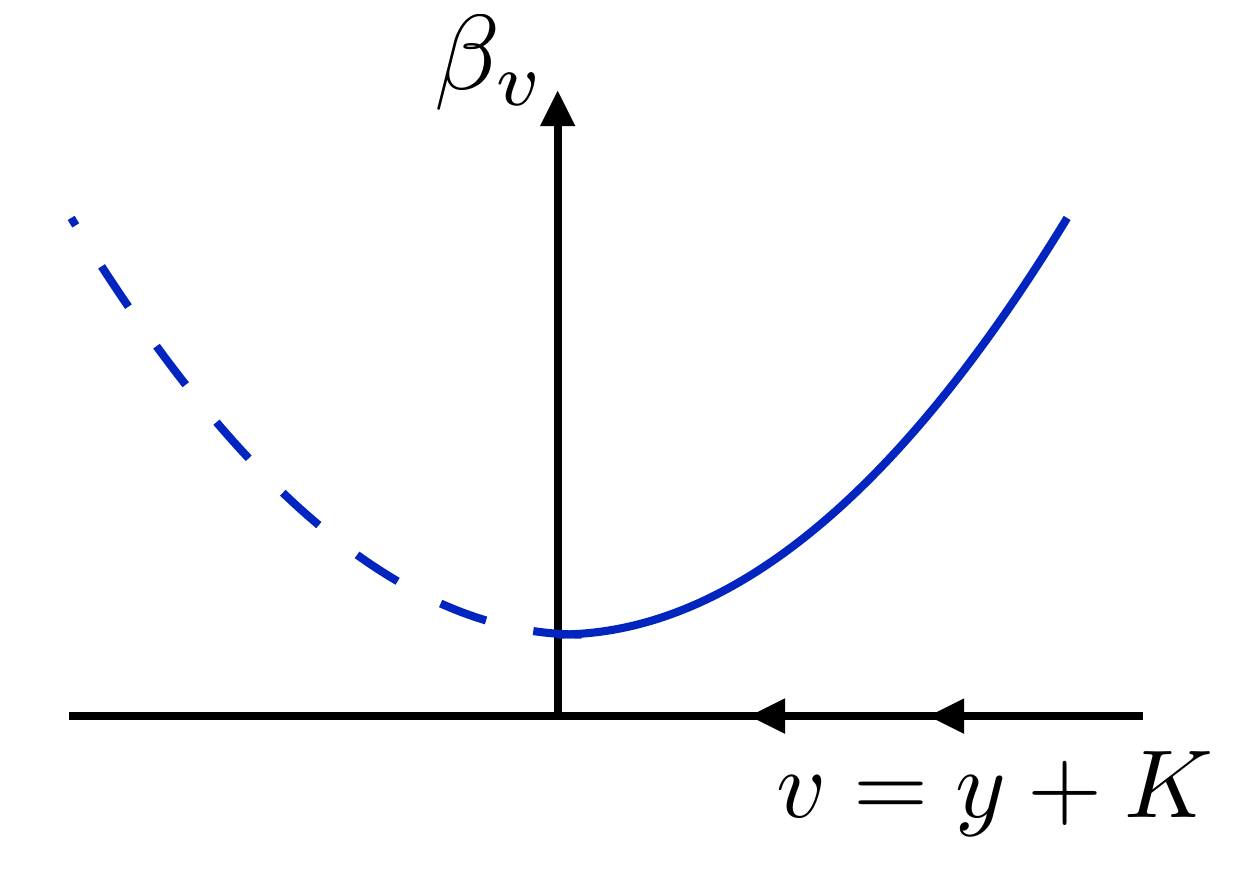} 
	\caption{{\it Left:} the RG flow diagram for the BKT transition. {\it Right:} the beta-function for the coupling combination $v$.}
	\label{fig:BKT}
\end{figure}

Let us go back to the general setup and consider the consequences. By shifting and rescaling $K$ we can assume that $K_c=0$ and $\Delta_1=d+K+O(K^2)$ near $K=0$. We perturb $\calT(0)$ by $K \calO_0+ y \calO_1$ and study the RG flow, which to the lowest order in $K,y$ has the form:
\beq
\beta_K=\frac 12 S_d C_{110} y^2,\qquad\beta_y=Ky\,.
\eeq
Notice that since $K$ is exactly marginal, its beta-function vanishes in absence of $y$ perturbation. Further rescaling the couplings to absorb the OPE coefficient, the new beta-functions take the simple form:
\beq
\beta_K= y^2,\qquad\beta_y=Ky\,.
\label{eq:betaKy}
\eeq
The RG flow diagram is in Fig.~\ref{fig:BKT}. Consider the flows starting at $K_0>0$ and at $y_0>0$. If $y_0<K_0$ then we end at a CFT, while for $y_0>K_0$ we flow `to the unknown' (presumably some gapped phase). 
When the microscopic theory is varied along the `micro' line, transition between the two regimes will happen at some non-universal value $K=K_*$. This is the BKT transition.

To study the flow to the unknown, we introduce two combinations of the couplings $u=y^2-K^2$ and $v=y+K$, in terms of which the RG equations take the form
\beq
\beta_u=0,\qquad \beta_v=\half v^2+\half u\,.
\eeq
The first equation means that $u$ stays constant in this one-loop approximation, while
in the second equation we recognize our walking beta-function equation \reef{eq:RGwalk} with $v$ and $u$ some trivial rescalings of $\lambda$ and $y$. Let us fix then $u>0$ and consider a flow of $v$ which starts at some positive $v=v_0\ll 1$ and heads towards $v=0$. Differently from the walking scenario \cite{Kaplan:2009kr}, the flow terminates at $v\sim u$, at which point $y-K=u/v=O(1)$ and we go out of the regime of validity of the original beta-functions \reef{eq:betaKy}. We get only one half of the walking RG trajectory, and the arising exponential hierarchy is given by the equation of the form \reef{eq:xiWalk} with an extra $\half$ in the exponent.

The main differences between walking and the BKT transition are as follows. 
One key difference is that the true couplings which control the weakness of the perturbation around $\calT(0)$ are $K$ and $y$, not their combinations $u$, $v$ which make the RG equations assume a simple form. That's why RG in the BKT transition breaks down for $v$ close to 0, while it's perfectly fine and weakly coupled in the walking RG running for $\lambda\sim 0$. The second difference is that the combination $u$ of the couplings, which enters as a fixed parameter into $\beta_v$, only remains constant in the one-loop approximation. This was not so for the walking beta-function \reef{eq:RGwalk}, where the analogous parameter $y$ was not renormalized to any order. 
 A deeper structural reason for the latter difference is that in the BKT transitions, the CFTs $\calT(K)$ all have the same symmetry and are related by an exactly marginal deformation (and hence parameter $K$ can flow). All of these CFTs, be that for $K<K_c$ or $K>K_c$, are unitary and nothing goes into the complex plane. On the contrary, in the considered examples of the walking scenario the family of CFTs all have a different global symmetry and are not related by an exactly marginal deformation.

Notice as well that the leading exponent in the BKT scaling is not universal since there is no universal relation between the parameters of the microscopic theory and the coupling $u$, while on the other hand, the leading exponent in \reef{eq:xiWalk} is universal, it depends only on the CFT data. For the above reasons, we propose to avoid calling hierarchy \reef{eq:xiWalk} `BKT scaling' when discussing the walking scenario. We propose to refer to it as the `walking scaling'.

\section{Further facts about the Potts model}
\label{sec:furtherPotts}

In this section we collect a few further facts about the Potts model which, while not central to our main line of reasoning, may turn out useful for non-experts.

\subsection{`Breakdown' of Landau-Ginzburg paradigm}

Historically, the first successful approach to phase transitions was the Landau-Ginzburg (LG) paradigm \cite{LL}. Although from modern perspective, limitations of this paradigm are well known, it remains a highly influential stepping stone in our thinking about the physics of phase transitions. The basic assumption of LG paradigm is that one can describe continuous phase transitions by considering the fluctuations of the order parameter. One considers an effective Lagrangian built out of the order parameter, which respects the same symmetries of the model and is supposed to describe the coarse-grained physics of it. In the original formulation, one applies the mean field approximation by neglecting fluctuations, and studies the order of the transition. For example, for the Ising model the order parameter is a scalar $\varphi$ odd under the $\mathbb{Z}_2$ symmetry. We end up with a Lagrangian given by even powers of $\varphi$, and this correctly predicts a second-order phase transition. For the Potts model with $Q\ge 3$, limiting ourselves to integer $Q$ for this discussion, the situation is different. The order parameter is the magnetization $\varphi_a$, a vector under $S_Q$, with $a=1,\ldots,Q$, and subject to the constraint $\sum_a \varphi_a=0$. The symmetry $S_Q$ acts by shuffling the indices $a$ around, and it is possible to construct a cubic term which is singlet under $S_Q$. Since all the terms not forbidden by the symmetry have to be included in our effective description, this term has to be considered. Within the original LG rules, the presence of the cubic term would imply that the phase transition is first-order for all $Q\ge 3$. That this prediction is not correct in the case of $Q=3,4$ and $d=2$ is a breakdown of the Landau-Ginzburg theory. In the case at hand the breakdown is usually explained by saying that the fluctuations of the order parameter are significant, and so what was first-order transition in mean field description becomes second-order in reality.

The words `LG description' are sometimes used in the theory of critical phenomena in a way different from the above \cite{Zamolodchikov:1986db}. Namely, one considers a UV-complete QFT  built out of scalar fields with relevant interactions which, for some value of the couplings, flows in the IR to a CFT of interest (the same CFT may describe the continuous phase transition of a lattice model). Such an LG description exists for all unitary minimal models \cite{Zamolodchikov:1986db}, for the Yang-Lee CFT $\calM_{2,5}$ \cite{Cardy:1985yy}, as well as for some other non-unitary minimal models \cite{Amoruso}. For 2d $Q=3,4$ Potts model, natural candidates for such LG descriptions are the $S_Q$-symmetric Lagrangians considered in \cite{Zia:1975ha,AmitPotts} which contain both cubic and quartic interaction terms.

\subsection{First-order phase transition at large $Q$}
\label{sec:app1st}

Consider the Potts model in $d$ dimensions with $Q\gg 1$. We will argue that the phase transition is first-order.  Consider first the zero-temperature ($v=\infty$) fully ordered state which in the cluster definition corresponds to the lattice-filling $X$, and the infinite-temperature ($v=0$) fully disordered state corresponding to the empty $X$.
The free energies per site of these two states are:
\beq
f_{\rm Ord} = d\log v, \quad f_{\rm Dis} =\log Q\,.
\eeq
Assuming that these states adequately describe physics all the way to the transition (which will be argued to be the case for $Q\gg 1$), we determine the approximate transition temperature by equating these two free energies: $v_c\approx Q^{1/d}$.\footnote{In $d=2$ this guess turns out to be exact for any $Q$, as follows from the self-duality of the model.} 
To show that this guess is self-consistent, we do the low-temperature expansion around the ordered state and the high-temperature expansion around the disordered state. Normally these expansions would converge only for very large and very small $v$ respectively. But for $Q\gg 1$ they actually converge all the way to the transition. At low expansion orders the smallness of corrections is easy to check. E.g.~the first correction to the disordered state comes from $X$ having one bond, and is suppressed by $v/Q$, which remains $\ll 1$ for $v\lesssim v_c$.
On the other hand the first few correction terms to the ordered state correspond to removing $k\le 2d-1$ bonds and are suppressed by $1/v^k$, which is $\ll 1$ throughout the region $v\gtrsim v_c$. At $k=2d$ we can finally create one more cluster---an isolated point. This gives a contribution $\sim Q/v^{2d}$, still suppressed. This can be made systematic by writing down the full $1/Q$ expansion series around the ordered and disordered state.\footnote{The $1/Q$ expansion in the Potts model was originally discussed in \cite{Ginsparg:1980ny,KOGUT1,KOGUT2,KIM,ParkKim}. See especially Eqs. (7), (8) in \cite{Bhattacharya:1996ix}.} This argument can be made mathematically rigorous using the Pirogov-Sinai theory, see \cite{Laanait1991} and \cite{grimmett2006random}, section 7.5.

To summarize, the fully ordered and fully disordered state survive, up to small correction, all the way to the transition temperature where they coexist. Both of these states clearly have $O(1)$ correlation length. Finiteness of the correlation length and phase coexistence mean that the transition is first-order at large $Q$.

\subsection{Generalization to $d>2$}
\label{sec:Pottsd>2}

We have seen that, in $d=2$, the order of the transition depends on the value of $Q$. It is believed that, in a general number of dimension $d$, the transition is second-order for $0<Q\le Q_c(d)$ and first-order for $Q>Q_c(d)$. As we have seen, $Q_c(2)=4$.\footnote{This has been proven rigorously \cite{Hugo1,Hugo2}.}  In three dimensions, it is known the transition is continuous for $Q=2$ and (weakly) first-order for $Q=3$.\footnote{See \cite{Wu} for the older evidence of a first-order transition in $d=3$, $Q=3$.} The value of $Q_c$ was found to be $Q_c(3)\approx 2.45$ in some Monte Carlo studies \cite{Kosterlitz}. For $d\ge 4$, it is known that $Q_c(d)=2$.

\begin{figure}
	\centering
	\includegraphics[scale=0.4]{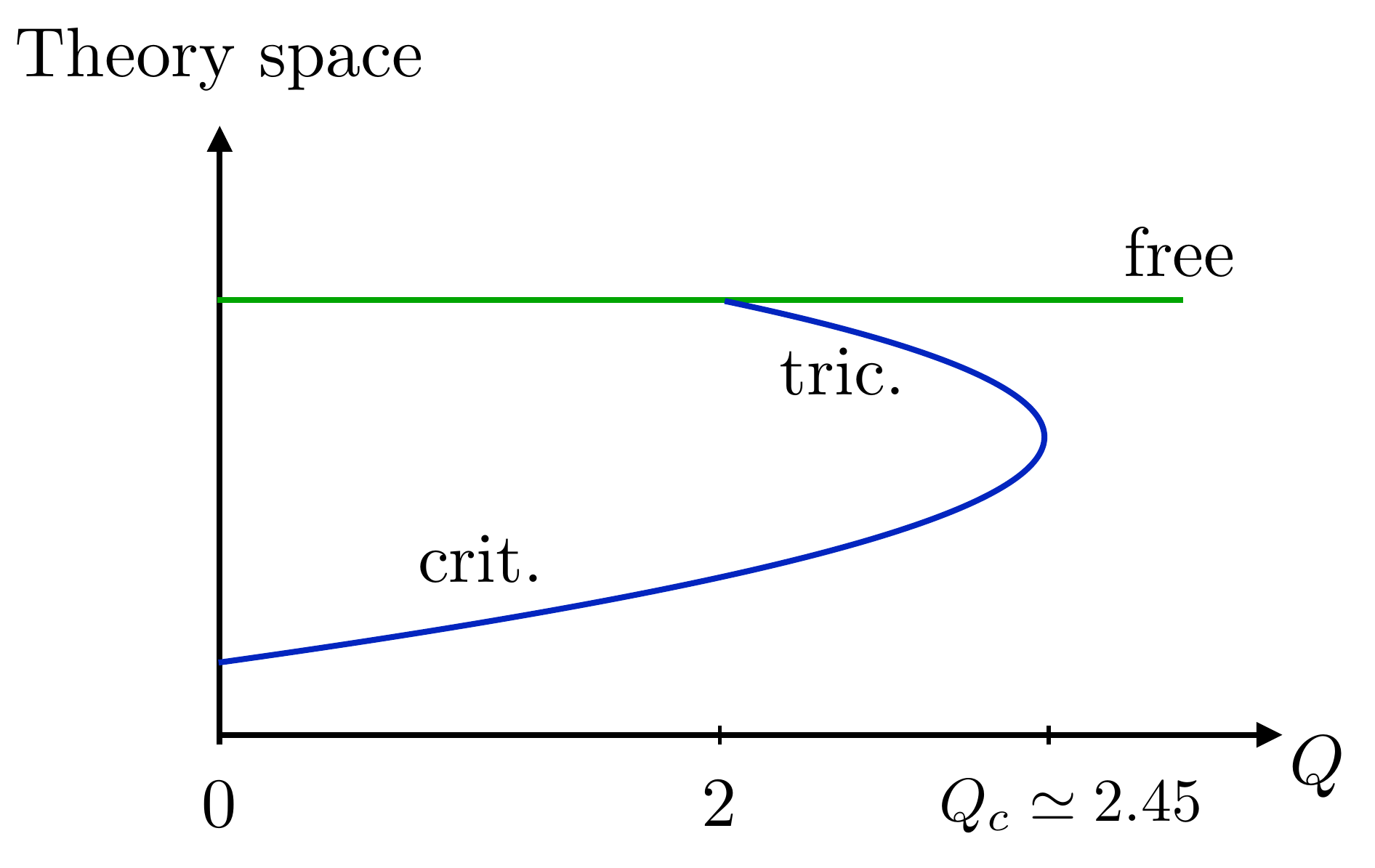}
	\caption{Critical and tricritical Potts model for $d=3$ as a function of $Q>0$. The two fixed points annihilate at $Q\approx 2.45$. At $Q=2$ the tricritical line intersects with the line of gaussian fixed points (see the text). }
	\label{fig:d=3}
\end{figure}

When it comes to the critical and tricritical fixed point annihilating, there is evidence from RG that it happens in $d\ne 2$ similarly to 2d \cite{Nienhuis1981,Newman:1984hy}. It seems thus reasonable to assume that in 3d the two fixed points annihilate at $Q=Q_c (3)\approx 2.45$.
One difference of $d>2$ from 2d is that there exist a value $Q_m(d)$ such that at $Q=Q_m(d)$ the line of tricritical fixed points meets the gaussian (free) line \cite{Nienhuis1981}.
In 3d, we expect $Q_m(3)=2$, in accord with the upper critical dimension for the Ising tricritical point being $d=3$.
See Fig.~\ref{fig:d=3} for the conjectured summary of the situation in $d=3$. 

For the $Q=3$ 3d Potts model the transition is weakly first-order, with the correlation length still largish, $\xi \sim 10$ \cite{JANKE1997679}. The complex CFT picture developed in our work may be relevant in this case. 

One could wonder if it's possible to start from a gaussian fixed point and vary the value of $Q$ in order to get a weakly coupled interacting theory, for example for the tricritical Potts model for $Q=2+\delta$, $d=3$ or the critical Potts model for $Q=2-\delta$, $d=4$, with $\delta $ small. In the latter context, this question was examined in \cite{Newman:1984hy}, and the answer is negative. It was found that, in $d=4$, the theory in the limit $Q\to 2^-$ reduces to two decoupled sector, one being a free theory describing the Ising model, and a strongly coupled second sector describing the Potts fields. While it is true that at $Q=2-\delta$ the two sectors interact weakly, the full theory is not perturbative. Using this framework, Ref.~\cite{Newman:1984hy} developed a theory describing the critical and tricritical fixed points merger in $d=4-\eps$ dimensions.

\section{Walking in large-$N$ theories}
\label{sec:largeN}
In this appendix we discuss an example of walking behavior in field theories with large-$N$ counting. As in the examples discussed in the main text we assume existence of two families of fixed points that depend on a parameter, $x$, and that merge for some critical value of this parameter, $x_c$. We also assume that at least when the parameter is close to its critical value there exists an RG flow connecting the fixed points. At large $N$, the corresponding flow was studied in \cite{Gubser:2002vv} by means of the Hubbard-Stratonovich transformation, where it was shown that it exists as long as one of the CFTs contains a double-trace operator which is weakly relevant. Here we give a simple description using conformal perturbation theory (CPT) that is valid in the vicinity of the merger point. Let us call the operator which triggers the flow $[\calO\calO]$,
and denote its dimension $d+\gamma_{\rm UV}$ at the UV fixed point and  $d+\gamma_{\rm IR}$ at the IR one.
Then $\gamma_{\rm UV}<0<\gamma_{\rm IR}$ and they go to zero when $x=x_c$. First of all, let us show that the operator responsible for the flow has to be a double-trace operator.\footnote{In theories with large $N_c$ and $N_f$ we will use single-traceness condition with respect to both $N_c$ and $N_f$.} To do this, recall the leading-order formula for the change in anomalous dimensions:
\beq
\gamma_{\rm IR}-\gamma_{\rm UV}=2 S_d g_{FP} C^{[\calO\calO]}_{[\calO\calO][\calO\calO]},
\eeq
where $g_{FP}$ is the value of the coupling constant at which the IR CFT is reached. If instead of $[\calO\calO]$ we tried to use some single-trace operator, say operator $\calO$ from which we are ``building" $[\calO\calO]$, its OPE coefficients of the form  $C^{\calO}_{\Phi \Phi}$, where $\Phi$ is any operator, including $\calO$ itself, would be suppressed by $1/N$. Correspondingly, $g_{FP}$ would have to be at least of order $N$ and the flow wouldn't be perturbative. Here we are assuming that at least some anomalous dimensions in two CFTs are different at the $O(1)$ order in $1/N$. Instead, the double-trace operator OPE coefficients $C^{[\calO\calO]}_{[\calO\calO][\calO\calO]}$ and $C^{[\calO\calO]}_{\calO \calO}$ are $O(1)$ and as long as $\gamma$'s are small we expect to be able to control the flow within CPT around the UV fixed point.

There is one simple cross-check that we can make at the leading order. Dimension of $\calO$ is given by
\beq
\dim(\calO)=\frac{d}{2}+\frac{1}{2}\gamma_{\rm UV(IR)}+O(1/N)
\eeq
and for consistency we need $C^{[\calO\calO]}_{[\calO\calO][\calO\calO]}=2C^{[\calO\calO]}_{\calO\calO}+O(1/N)$, so that $\calO$ gets the right dimension for the same value of $g$. Since to leading order the OPE coefficients can be calculated by Wick contractions it is easy to check that this relation indeed holds for canonically normalized operators.
\begin{figure}[h]
	\centering
	\includegraphics[scale=.4]{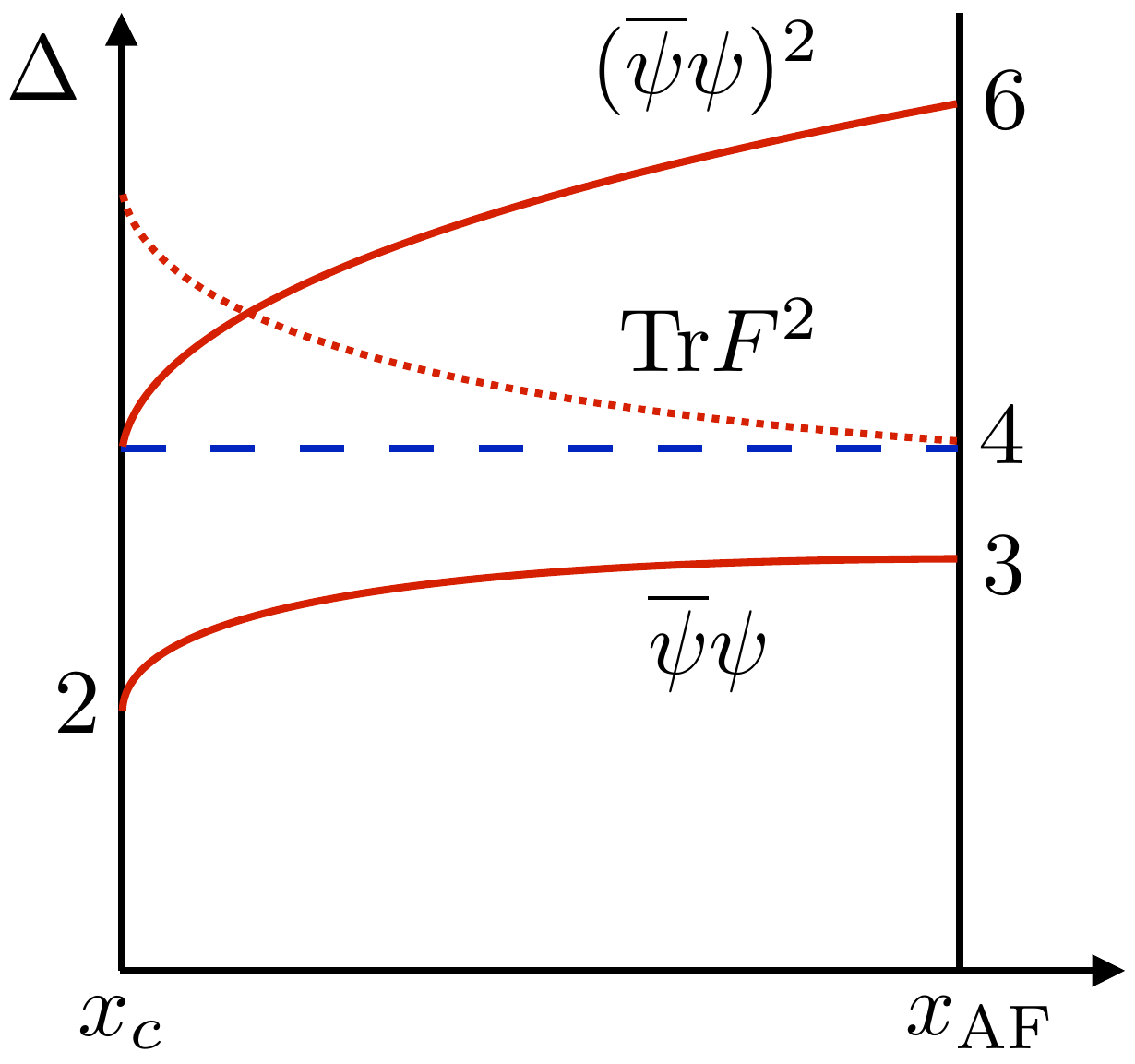}
	\caption{Schematic behavior of the operator dimensions at the BZ fixed point as a function of $x=N_f/N_c$, at $N_c=\infty$. The dimension of $(\bar \psi\psi)^2$ is twice that of $\bar\psi\psi$. {All dimensions have a square-root singularity at $x=x_c$.}} 
	\label{fig-dimsBZ}
\end{figure}

{Arguments above simply relied on some sort of $1/N$ expansion. In particular, they apply to gauge theory in the large $N_c$,$N_f$ limit holding $x=N_f/N_c$ fixed. This is the large $N$ limit of the Banks-Zaks-like theories discussed in section \ref{sec:BZ}. In this context we arrive at the following conclusion.} For $x=x_{\rm AF}$ when the BZ fixed point is free all operators with low dimensions can be easily classified. If we are looking for an operator that for $x=x_c$ becomes marginal and controls the walking behavior for $x<x_c$ then at large $N$ this operator must be a double-trace singlet. As it was advocated in \cite{Kaplan:2009kr}, good candidates are four-fermion operators which for $x=x_{\rm AF}$ have dimension 6.

If this picture is right, the schematic behavior of operator dimensions at the BZ fixed point in $d=4$ in the range $x_c<x<x_{\rm AF}$ has, in the strict $N_c=\infty$ limit, schematic form shown in Fig.~\ref{fig-dimsBZ}. Since $(\bar\psi \psi)^2$ starts at dimension 6 and is expected to become marginal at $x=x_c$, there should be a level crossing between this operator and ${\rm tr} F^2$. An alternative picture in which it's ${\rm tr} F^2$ becomes marginal is, as we said, disfavored because the three-point function of this single-trace operator vanishes at $N_c=\infty$, and so it's unsuitable for generating a flow from QCD${}^*$ to BZ with expected properties.

Of course at finite but large $N_c$ we expect that level crossing in Fig.~\ref{fig-dimsBZ} will be resolved by a small amount. In this case, the operator which becomes marginal at $x=x_c$ is continuously connected to ${\rm tr} F^2$, but it still makes sense to label it as double trace $(\bar\psi \psi)^2$ because its properties are similar to those of the latter in $N_c=\infty$ limit.


\small

\bibliography{../../../1st-Biblio}
\bibliographystyle{utphys}

\end{document}